\documentclass[aps,prb,showpacs,preprintnumbers,twocolumn,superscriptaddress,nofootinbib]{revtex4-2}
\usepackage{amsmath,amssymb}
\usepackage{bm}
\usepackage{tipa}
\usepackage{upgreek}
\usepackage{comment}
\usepackage{mathrsfs}
\usepackage{graphicx}
\usepackage{braket}
\usepackage{enumitem}
\usepackage{mathbbol}
\usepackage{booktabs}
\usepackage{gensymb}
\usepackage[normalem]{ulem}
\usepackage{color}
\usepackage[colorlinks,bookmarks=true,citecolor=blue,linkcolor=red,urlcolor=blue]{hyperref}
\usepackage{hyperref}

\usepackage{dsfont}
\usepackage{tikz-feynman,contour}

\def \k{{\mathbf k}}
\def \q{{\mathbf q}}

\def \r{{\mathbf r}}
\def \Q{{\mathbf Q}}

\def \tk{{\Tilde{\mathbf k}}}

\usepackage{listings}

\begin{document}
	
	\title{Electrons interacting with Goldstone modes and the rotating frame}
 \author{Konstantinos Vasiliou}
	\affiliation{Rudolf Peierls Centre for Theoretical Physics, Parks Road, Oxford, OX1 3PU, UK}
	\author{Yuchi He}
	\affiliation{Rudolf Peierls Centre for Theoretical Physics, Parks Road, Oxford, OX1 3PU, UK}
	\author{Nick Bultinck}
	\affiliation{Rudolf Peierls Centre for Theoretical Physics, Parks Road, Oxford, OX1 3PU, UK}
	\affiliation{Department of Physics, Ghent University, Krijgslaan 281, 9000 Gent, Belgium}

	\begin{abstract}
   We consider electronic systems with a spontaneously broken continuous symmetry. The scattering vertex between electrons and Goldstone modes is calculated over the entire Brillouin zone using the random phase approximation. This calculation reveals two things: (1) electrons always couple to both $\phi$ and $\partial_t\phi$, where $\phi$ is the Goldstone field, and (2) quasi-particles in a state with continuous symmetry breaking have to be defined in a rotating frame, which locally follows the fluctuations of the order parameter. The implications of these findings for electron spectral functions in both symmetry-broken and thermally disordered systems are discussed, and the examples of anti-ferromagnetism in the Hubbard model and spin spiral order in the three-band model are worked out in detail. 
	\end{abstract}
	
 	\maketitle

\tableofcontents
 	 
\section{Introduction}

Many strongly-correlated materials display continuous symmetry breaking in some part of their phase diagram. Some paradigmatic examples are spin-density wave orders in the cuprate and iron-based superconductors, and valley/spin orders in 2D moir\'e materials. 

As a result of the continuous symmetry-breaking, the low-energy spectrum contains gapless Goldstone modes. In this work we set out to calculate the scattering vertex between these Goldstone modes and the electrons. Specifically, our goal is to obtain an expression for the scattering vertex which is valid over the entire Brillouin zone, and not just in the long-wavelength limit for the Goldstone modes. One reason for going beyond a long-wavelength approximation is that, by analogy with acoustic phonons, Goldstone modes of the electron liquid are expected to decouple from the electrons at long wavelengths. Hence one expects the most important scattering processes to be those where an electron emits or absorbs a zone-boundary Goldstone mode. 

The starting point of our analysis is a simple mean-field description of the broken-symmetry state. From the optimal broken-symmetry Slater determinant, we then construct a path-integral representation of the partition function in order to study fluctuations on top of the mean-field state. Via the random phase approximation (RPA)\footnote{Note that in this work RPA means summing both bubble and ladder diagrams, such that is a conserving approximation in the sense of Baym and Kadanoff \cite{Baym1961}. It is also equivalent to applying the time-dependent variational principle \cite{Goldstone1959,Kramer}, and hence is not a perturbative calculation which is only applicable to weakly-interacting systems.} we obtain both the Bethe-Salpeter equation for the collective modes (whose solution gives the Goldstone mode energies and wavefunctions), and an explicit expression for the electron-Goldstone mode scattering vertex. Using these results, we then construct an effective electron-boson model with both fermionic and bosonic fields, which emulates the interacting electron-Goldstone mode system. Towards the end of the paper we illustrate our formalism with two examples: anti-ferromagnetism in the Hubbard model, and circular spin-spiral order in the three-band model.

\subsection{Summary of results and connection to previous works}

Even though both the goal (obtaining the electron-Goldstone mode scattering vertex) and the methods (mean-field theory and RPA) of this work are straightforward and have previously been discussed in classic works such as e.g. Refs. \cite{Bickers1989,Schrieffer1989,Chubukov1992}, we nevertheless find that obtaining a result which is in line with physical expectations requires some non-trivial (and to the best of our knowledge new) steps. In this section we briefly discuss what these non-trivial steps are, why they are necessary, and what interesting physics is hiding behind them.

Via the RPA analysis we obtain a scattering vertex which in the long-wavelength limit agrees with the previous result of Ref.~\cite{Watanabe2014}. Ref.~\cite{Watanabe2014} exclusively focused on intra-band scattering processes, whereas in this work we will also consider inter-band scattering. As was shown in Ref.~\cite{Watanabe2014}, intra-band scattering by Goldstone modes vanishes in the the long wavelength limit if the associated broken symmetry generators commute with translations. However, the expression for the scattering vertex obtained from our RPA analysis, and thus also the vertex obtained in Ref. \cite{Watanabe2014}, implies that long-wavelength Goldstone modes do not completely decouple from the mean-field electrons\footnote{Mean-field electrons = electrons occupying single-particle eigenstates of the mean-field Hamiltonian.}. Instead, they give rise to inter-band scattering with a scattering vertex of order $\Delta$, the mean-field bandgap which results from the spontaneous symmetry breaking. As a result, the factor $\Delta^{-1}$, which suppresses highly non-resonant processes where a low-energy Goldstone mode scatters an electron across the bandgap, is exactly cancelled by the strength of the vertex. Inter-band scattering by long-wavelength Goldstone modes will thus produce non-negligible self-energy corrections for the mean-field electrons, making it unclear whether they can be used to describe the quasi-particles of the system.  

However, we observe that the spurious inter-band contribution to the electron-Goldstone mode vertex is `pure gauge', meaning that it can be removed by a gauge transformation in the path integral. This gauge transformation implements a spatially-dependent symmetry transformation on the mean-field electrons. In particular, the transformed electron fields are defined in a rotating frame which follows the local fluctuations of the order parameter. Crucially, electrons in the rotating frame completely decouple from long-wavelength Goldstone modes. We note that a similar rotating frame has already been introduced in previous studies of superconducting \cite{Ramakrishnan1989,Paramekanti2000} and magnetic \cite{Shraiman1988,Shraiman1990,Schulz1990,Schulz1994,Weng} orders, but for different reasons. In particular, these previous works did not obtain the Goldstone mode energies and wavefunctions by solving the Bethe-Salpeter equation, and did not motivate the rotating frame by requiring a decoupling between electrons and Goldstone modes at long wavelengths.

The mean-field electrons are simply frequency-independent linear combinations of the bare electrons, and hence their correlation functions are directly related to response functions measured in experiment. But we argued above that the mean-field electrons are not a good approximation to the true quasi-particles of the system -- one should instead use the electrons in the rotating frame to approximate the quasi-particles. However, the electron propagator can be approximated as a convolution (in frequency and momentum space) of two propagators of well-defined quasi-particles: fermions in the rotating frame, and the Goldstone modes (we momentarily ignore the possibility of Landau damping for the Goldstone modes). This is explained in detail in Sec. \ref{sec:spectfun}. This result has previously also been obtained in Ref. \cite{Borejsza2004}, where, similarly to Ref. \cite{Schulz1990}, the rotating frame is introduced to correctly describe fluctuations beyond mean-field theory via a Hubbard-Stratanovich decoupling of the interaction. Again, this work did not motivate the necessity of a rotating frame by deriving the electron-Goldstone mode scattering vertex in the RPA formalism as we do here. We note that a similar expression for the fermion two-point function as a convolution of two propagators is also obtained in theories where the electron is assumed to fractionalize in a spinon and a holon \cite{Florens2004,Qi2010,Tang2013,Chatterjee2017,Scheurer2018,Sachdev2019,Bonetti2022}. 

The RPA approach describes small fluctuations around the mean-field state. In the effective electron-boson model, which we construct to emulate the interacting electrons and Goldstone modes at the RPA level, the Goldstone fields are a collection of real scalar fields. However, once the dispersion relation of small order parameter fluctuations is known, it is straightforward to improve the theory and incorporate the correct global structure of the order parameter manifold by describing the Goldstone mode dynamics with a non-linear sigma model. This is especially important in 2D, where the non-linear sigma model is disordered at any non-zero temperature, and hence ensures that the effective electron-boson model respects the Hohenberg-Mermin-Wagner theorem.

As a final point, let us elaborate on the physical behaviour which can be extracted from the expression for the microscopic fermion propagator as a convolution of the Goldstone propagator and the propagator of the fermions in the rotating frame. In particular, let us consider 2D systems, and focus on temperatures $T$ which are much smaller than the mean-field bandgap. If we use the correct non-linear sigma model action for the Goldstone mode dynamics, the Goldstone mode propagator at finite $T$ will be invariant under global symmetry transformations, and acquire a thermal mass. As a result, the propagator for the microscopic fermions (which is a convolution containing the Goldstone mode propagator) will also be symmetric (this is explained in detail in Sec. \ref{sec:spectfun}, and illustrated via the examples in Sec. \ref{sec:ex1} and \ref{sec:ex2}). However, the spectral weight of the microscopic electrons is still predominantly determined by the poles of the fermion propagator in the rotating frame. As we assume $T$ to be much smaller than the mean-field bandgap, the poles of the rotating-frame fermion propagator will be located at the same energies of the symmetry-broken mean-field band spectrum. As a result, we find that even though the microscopic fermion propagator is invariant under the continuous global symmetry, it nevertheless produces a spectral weight that is predominantly determined by the spectrum of the broken-symmetry state -- and hence is very different from the spectral weight of the non-interacting fermions. We thus see that the introduction of a rotating frame, which is necessary to have the electrons decouple from the long-wavelength Goldstone modes, automatically leads to an expression for the electron propagator which captures the physically intuitive behaviour of an electron system with an order parameter that has well-defined magnitude, but whose orientation is disordered by thermal fluctuations.

\subsection{Structure of the paper}

The remainder of this work is organized as follows. We start in Sec. \ref{sec:MF} by introducing the mean-field starting point of our analysis. In Sec. \ref{sec:pathint} we use the optimal Slater determinant, which we obtain from the mean-field analysis, to construct a path integral representation of the partition function, which can be used to study fluctuations beyond mean-field. Using this path integral, we introduce the infinite series of RPA Feynman diagrams that give rise to the Bethe-Salpeter equation, and also the effective interaction between electrons mediated by collective mode exchange, in Sec. \ref{sec:Veff}. In Sec. \ref{sec:interacting} we first discuss the general properties of the Goldstone mode wavefunctions that we obtain by solving the Bethe-Salpeter equation. We then construct an effective electron-boson model which mimics the interacting electron-Goldstone mode system at tree-level. In Sec. \ref{sec:rotframe} we use the properties of the Goldstone mode wavefunctions to study the electron-Goldstone mode scattering vertex obtained from the RPA analysis in the long-wavelength limit. Here we uncover the problematic inter-band scattering processes which do not vanish in the long-wavelength limit. In the same section, we then explain how this problem can be solved by doing a gauge transformation in the path integral, and defining fermions in a rotating frame. The implications of the rotating frame for the electron spectral functions are discussed at the end of this section. Finally, we illustrate the general formalism for anti-ferromagnetism in the Hubbard model in Sec. \ref{sec:ex1}, and for circular spin-spiral order in the three-band model in Sec. \ref{sec:ex2}. The appendix reviews some properties of the Bethe-Salpeter equation and its solutions.

\section{Mean-field starting point}\label{sec:MF}

In this section we set the stage for our main results presented below. In particular, we introduce the mean-field theory which is the starting point for our calculations. All the concepts discussed here are standard, and the main purpose of this section is therefore to introduce the context and notation necessary to understand the following sections.

We consider interacting electron systems described by the following general Hamiltonian:

\begin{equation}\label{Hrealspace}
H = \sum_{\r,\r'}\sum_{a,b} h(\r-\r')_{ab} c^\dagger_{\r,a} c_{\r',b} + \frac{1}{2}\sum_{\r,\r'}V(\r-\r'):n_{\r} n_{\r'}:\,,
\end{equation}
where $\r$ denote the sites of a Bravais lattice in $d$ spatial dimensions, $h(\r)$ is a general hopping Hamiltonian, and $n_\r = \sum_a c^\dagger_{\r,a}c_{\r,a}$ is the electron density at site $\r$. We also assume that $h(\r)$ is short range, and $V(\r)$ has a well-defined Fourier transform (for example it could be a screened Coulomb potential). Our results can be generalized to other Hamiltonians, for example with other (not density-density) interactions, but for ease of presentation we do not consider these generalizations here.

We will make use of the translation invariance and work in momentum space, where the Hamiltonian in Eq. \eqref{Hrealspace} takes the form
\begin{equation}
    H = \sum_\k \sum_{a,b} h(\k)_{ab}c^\dagger_{\k,a}c_{\k,b} + \frac{1}{2}\sum_\q V(\q) :n_{\q} n_{-\q}:\,,
\end{equation}
where
\begin{equation}
    n_\q = \frac{1}{\sqrt{N}} \sum_\k \sum_a c^\dagger_{\k+\q,a}c_{\k,a}\,,
\end{equation}
with $N$ the number of sites, is the Fourier transform of the electron density.

We are interested in Hamiltonians with a continuous symmetry, such as spin rotation symmetry, which gets broken spontaneously. The starting point of our analysis is a mean-field treatment of the symmetry breaking at zero temperature. In particular, we assume that the Slater determinants which minimize the energy of $H$ are not invariant under the continuous symmetry of the Hamiltonian. Let us choose one such Slater determinant and write it as

\begin{equation}\label{SlaterDet}
    |\psi_0\rangle = \prod_{\k}\prod_{i=1}^{N_{o}(\k)}c^\dagger_{\k,i}|0\rangle\,,
\end{equation}
where $c^\dagger_{\k,i}$ creates an electron in a mean-field single-particle state, and $N_o(\k)$ is the number of occupied states at momentum $\k$. Note that we have assumed that the optimal Slater determinants preserve the translation symmetry. Also this restriction can be relaxed, but we will not do this here. The mean-field single-particle states (both occupied and unoccupied) will be labeled by Greek indices, and the corresponding creation operators are defined as
\begin{equation}\label{eq:basistransform}
    c^\dagger_{\k,\alpha} = \sum_{a} u_\alpha^a(\k) c^\dagger_{\k,a}\,,
\end{equation}
where $|u_\alpha(\k)\rangle$ are a set of orthonormal vectors at every $\k$: $\langle u_\alpha(\k)|u_\beta(\k)\rangle = \delta_{\alpha\beta}$. We will use the convention that $i,j,k$ denote occupied mean-field states, whereas $m,n,o$ denote the unoccupied states. The single-particle correlation matrix of the optimal Slater determinant in Eq. \eqref{SlaterDet}, which by assumption is diagonal in momentum space, is thus given by
\begin{eqnarray}
    P(\k)_{\alpha\beta} & := & \langle\psi_0|c^\dagger_{\k,\beta}c_{\k,\alpha}|\psi_0\rangle=\delta_{\alpha\beta}n_\alpha(\k)\,,
\end{eqnarray}
with $n_\alpha(\k)\in\{0,1\}$, $n_{i/j/k}(\k)=1$ and $n_{m/n/o}(\k)=0$.

Next we construct the Hartree-Fock mean-field Hamiltonian using $P(\k)$. For future use we find it convenient to first rewrite the exact Hamiltonian in the mean-field basis:
\begin{eqnarray}
    H & = & \sum_\k h(\k)_{\alpha\beta} c^\dagger_{\k,\alpha} c_{\k,\beta} + \frac{1}{2}\sum_\q V(\q) :n_\q n_{-\q}:\\
    n_\q & = & \frac{1}{\sqrt{N_s}}\sum_\k \sum_{\alpha,\beta}c^\dagger_{\k-\q,\alpha}[\Lambda_\q(\k)]_{\alpha\beta}c_{\k,\beta}\,,
\end{eqnarray}
where $h(\k)_{\alpha\beta} = \sum_{a,b}u^{a*}_\alpha(\k)h(\k)_{ab}u^b_\beta(\k)$, and
\begin{equation}
    [\Lambda_\q(\k)]_{\alpha\beta} := \langle u_\alpha(\k-\q)|u_\beta(\k)\rangle\, .
\end{equation}
As we are only considering translationally invariant states, the Hartree Hamiltonian is trivial and is simply given by
\begin{equation}
    H_{H}[P] = V(0)\frac{N_e}{N} \sum_\k \sum_{\alpha}c^\dagger_{\k,\alpha} c_{\k,\alpha}\,,
\end{equation}
where $N_e$ is the number of electrons in the system. The Fock Hamiltonian is given by
\begin{equation}
    H_F[P] = \frac{-1}{N}\sum_{\q,\k}\sum_{\alpha,\beta,i}V(\q)  [\Lambda_{-\q}(\k-\q)]_{\alpha i}[\Lambda_{\q}(\k)]_{i \beta}c^\dagger_{\k,\alpha}c_{\k,\beta}
\end{equation}
The assumption that the Slater determinant in Eq. \eqref{SlaterDet} is a variational energy minimum is equivalent to the statement that the Hartree-Fock self-consistency equation is satisfied, which in our notation takes the following form:

\begin{equation}\label{selfconst}
\sum_{\alpha,\beta}h(\k)_{\alpha\beta}c^\dagger_{\k,\alpha}c_{\k,\beta} + H_H[P] + H_F[P] = \sum_{\alpha} E_{\k,\alpha}c^\dagger_{\k,\alpha}c_{\k,\alpha} \,,
\end{equation}
where $E_{\k,\alpha}$ are the mean-field single-particle energies satisfying $\text{sgn}(E_{\k,\alpha}) = 1-2n_\alpha(\k)$.

\section{Hartree-Fock path integral}\label{sec:pathint}

We are interested in fluctuations beyond mean-field theory, and in particular in the role of the Goldstone modes associated with the spontaneous symmetry breaking. To study these fluctuations, we will use the path integral formalism. However, we will not use the standard path integral construction which relies on fermionic coherent states constructed on top of the Fock vacuum. Instead, we will construct coherent states on top of the Slater determinant in Eq. \eqref{SlaterDet}. Working in the single-particle mean-field basis, coherent states associated with unoccupied states take on the conventional form:

\begin{eqnarray}
    |\psi_{\k,n}\rangle & = & (1-\psi_{\k,n}c^\dagger_{\k,n})|0\rangle \\
    \langle \bar{\psi}_{\k,n}| & = & \langle 0|(1 - c_{\k,n}\bar{\psi}_{\k,n})\,,
\end{eqnarray}
where $\psi_{\k,n}$ and $\bar{\psi}_{\k,n}$ are Grassmann numbers. These coherent states have the usual properties:

\begin{eqnarray}
    c_{\k,n}|\psi_{\k,n}\rangle & = & \psi_{\k,n}|\psi_{\k,n}\rangle\\
    \langle \bar{\psi}_{\k,n}|\psi_{\k,n}\rangle & = & e^{\bar{\psi}_{\k,n}\psi_{\k,n}}\, .
\end{eqnarray}
For the occupied states, we will use \emph{particle-hole transformed} coherent states, which we define as

\begin{eqnarray}
     |\bar{\psi}_{\k,i}\rangle & = & (1-\bar{\psi}_{\k,i}c_{\k,i})c^\dagger_{\k,i}|0\rangle \\
     \langle \psi_{\k,i}| & = & \langle 0|c_{\k,i}(1-c^\dagger_{\k,i}\psi_{\k,i})\,,
\end{eqnarray}
where $\bar{\psi}_{\k,i}$ and $\psi_{\k,i}$ are again Grassmann numbers. The particle-hole transformed coherent states satisfy following properties:

\begin{eqnarray}
    c^\dagger_{\k,i}|\bar{\psi}_{\k,i}\rangle & = & \bar{\psi}_{\k,i}|\bar{\psi}_{\k,i}\rangle\\
    \langle \psi_{\k,i}|\bar{\psi}_{\k,i}\rangle & = & e^{\psi_{\k,i}\bar{\psi}_{\k,i}}\, ,
\end{eqnarray}
 To construct a path integral representation of the partition function, we need to insert resolutions of the identity in terms of the coherent states, which are given by
 
\begin{eqnarray}
\int \mathrm{d}\psi_{\k,n}\int \mathrm{d}\bar{\psi}_{\k,n}e^{-\bar{\psi}_{\k,n}\psi_{\k,n}}|\psi_{\k,n}\rangle\langle \bar{\psi}_{\k,n}| & = & \mathds{1} \\
\int \mathrm{d}\bar{\psi}_{\k,i}\int \mathrm{d}\psi_{\k,i}e^{-\psi_{\k,i}\bar{\psi}_{\k,i}}|\bar{\psi}_{\k,i}\rangle\langle \psi_{\k,i}| & = & \mathds{1}
\end{eqnarray}
Using the above resolutions of the identity, the partition function at temperature $T$ can be written as

\begin{eqnarray}
    Z(T) & = & \text{tr}\left(e^{-H/T} \right)\\
    & = & e^{-E_0^{HF}/T}\int [D\psi]\int[D\bar{\psi}]\,e^{-S}\,, \label{pathint}
\end{eqnarray}
where $E_0^{HF} = \langle\psi_0|H|\psi_0\rangle$ is the Hartree-Fock variational ground state energy, and the action $S$ is given by
\begin{equation}\label{action}
\begin{split}
S = &\int_0^{1/T}\mathrm{d}\tau\, \sum_\k \sum_\alpha \bar{\psi}_{\k,\alpha} (\partial_\tau + E_{\k,\alpha})\psi_{\k,\alpha} \\
+\frac{1}{2N}&\sum_{\q,\k,\k'}V(\q) \left(\bar{\psi}_{\k-\q}\Lambda_{\q}(\k)\psi_{\k}\right)\left( \bar{\psi}_{\k'+\q}\Lambda_{-\q}(\k') \psi_{\k'}\right) \,,
\end{split}
\end{equation}
where in the last line we have suppressed the indices $\alpha,\beta,\dots$. A few comments are in order. First, note that the kinetic term of the action contains the mean-field single-particle energies, and not the bare band energies (which would correspond to the eigenvalues of $h(\k)$). Secondly, we emphasize that no approximation is involved in our derivation of the path integral -- Eqs. \eqref{pathint}, \eqref{action} constitute an exact representation of the partition function. Thirdly, the unusual form of the action, and the additional factor $\exp(-E_0^{HF}/T)$ in Eq. \eqref{pathint}, result from the different form of normal ordering required by the use of particle-hole transformed coherent states: one has to normal order the Hamiltonian with respect to the Hartree-Fock ground state, and not with respect to the Fock vacuum state. This different choice of normal ordering produces additional quadratic terms, which exactly correspond to the Hartree and Fock Hamiltonians, which combined with the bare kinetic term produce the mean-field single-particle energies via the Hartree-Fock self-consistency equation \eqref{selfconst}.

In working with the path integral in Eqs. \eqref{pathint}, \eqref{action} the conventional imaginary-time Feynman rules can be used, except for equal-time diagrams. These diagrams are usually defined by inserting a factor $e^{i\epsilon\omega_n}$, where $\omega_n$ is the fermionic Matsubara frequency, and taking the limit $\epsilon\rightarrow 0$ at the end of the calculation. Here, to reflect the different normal ordering, one has to add the factor $e^{i\epsilon\omega_n\text{sgn}(E_{\k,\alpha})}$ to the equal-time propagators. This automatically ensures that the Hartree and Fock self-energy diagrams vanish at zero temperature. 

To conclude this section, we present the diagrammatic Feynman rules that will be used in this work. First, the electron propagator is represented in the usual was as a straight line with an arrow:

\begin{center}
\begin{tikzpicture}
   \begin{feynman}
   \vertex (a) at (-0.3,0);
   \vertex (b) at (1,0);
   \vertex (c) at (2.7,0) {$=(i\omega_n - E_{\k,\alpha})^{-1}$\, .};

    \diagram* {
      (a) -- [fermion] (b),
    };
  \end{feynman}
\end{tikzpicture}
\end{center}
The interaction will be represented as a dashed line:

\begin{center}
\begin{tikzpicture}
   \begin{feynman}
   \vertex (a) at (0,-1) {$\k,\alpha$};
   \vertex (b) at (0,1) {$\k-\q,\beta$};
   \vertex (c) at (0.5,0);
   \vertex (d) at (2,0);
   \vertex (e) at (2.5,-1) {$\k'-\q,\sigma$};
   \vertex (f) at (2.5,1) {$\k',\lambda$};
   \vertex (g) at (3,0) {$=$};
   \vertex (h) at (5.8,0) {$ -V(\q)\big[\Lambda_{\q}(\k) \big]_{\beta\alpha} \big[\Lambda_{\q}^\dagger(\k') \big]_{\lambda\sigma}$\,.};

    \diagram* {
      (a) -- [fermion] (c),
      (c) -- [fermion] (b),
      (c) -- [scalar] (d),
      (e) -- [fermion] (d),
      (d) -- [fermion] (f),
    };
  \end{feynman}
\end{tikzpicture}
\end{center}
With these definitions and conventions in place, we can now turn our attention to the RPA fluctuations on top of the mean-field result, which is the topic of the next section.

\section{Effective RPA interaction and Bethe-Salpeter equation}\label{sec:Veff}

\begin{figure}
    \includegraphics[scale=0.61]{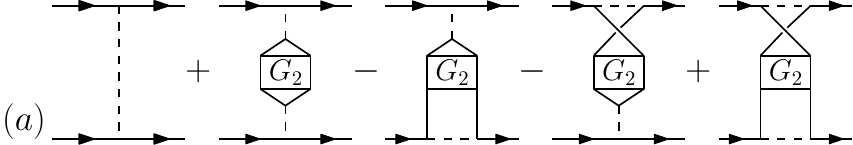}\\
    \vspace{0.7 cm}
    \includegraphics[scale=0.65]{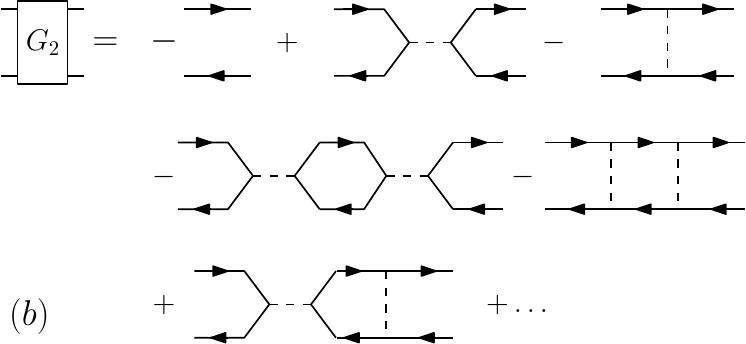}
    \caption{Definition of the effective interaction obtained from summing RPA diagrams.}
    \label{fig:Veff}
\end{figure}

In the previous section we introduced the Hartree-Fock path integral, which contains the bare Coulomb interaction. We now proceed by studying the effective interaction which results from summing all RPA diagrams as shown in Fig. \ref{fig:Veff} which contain, among others, the familiar polarization and Berk-Schrieffer diagrams \cite{Berk1966}. An important caveat of our RPA approach is of course that it ignores vertex corrections, which have been emphasized to play an important role in e.g. the two-particle self-consistent approach of Ref.~\cite{Vilk1997}. 

\begin{figure}
    \centering
    \includegraphics[scale=0.67]{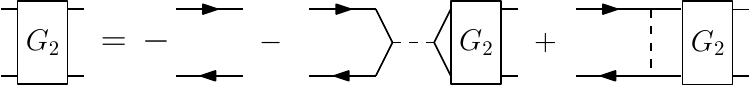}
    \caption{Diagrammatic representation of the Bethe-Salpeter equation for $G_2(\q,i\nu)$, the solution of which represents the infinite sum of RPA diagrams shown in Fig. \ref{fig:Veff}(b).}
    \label{fig:BetheSalpeter}
\end{figure}

The central object defined in Fig. \ref{fig:Veff} is called $G_2(\q,i\nu)$, which is a matrix diagonal in momentum $\q$ and bosonic Matsubara frequency $i\nu$. Writing out the indices explicitly, this matrix is $[G_2(\q,i\nu)]^{\k\alpha\beta}_{\k'\lambda\sigma}$. It is defined diagramatically in Fig. \ref{fig:Veff}(b) as the infinite sum of direct and exchange diagrams concatenated with electron propagators. An explicit expression for this infinite sum can be obtained by solving the familiar Bethe-Salpeter equation, which is shown diagramatically in Fig. \ref{fig:BetheSalpeter}. Written out explicitly, the Bethe-Salpeter equation for $G_2(\q,i\nu)$ is

\begin{widetext}
\begin{equation}
\begin{split}
    [G_2(\q,i\nu)]^{\k\alpha\beta}_{\k'\lambda\sigma} &=  \left[-T\sum_{i\omega_n}\frac{1}{i\omega_n - E_{\k,\alpha}} \frac{1}{i(\omega_n-\nu) - E_{\k+\q,\beta} } \right]\times \\
\bigg[\delta_{\alpha\lambda}\delta_{\beta\sigma}\delta_{\k,\k'} -\frac{1}{N_s}\sum_{\k''\mu\nu}\bigg( V(\q)&[\Lambda_\q(\k)]_{\beta\alpha}[\Lambda_{-\q}(\k''-\q)]_{\mu\nu} - V(\k''-\k')[\Lambda_{\k-\k''}(\k)]_{\mu\alpha}[\Lambda_{\k''-\k}(\k''-\q)]_{\beta\nu}\bigg)[G_2(\q,i\nu)]^{\k''\mu\nu}_{\k'\lambda\sigma}\bigg]
\end{split}
 \end{equation}
\end{widetext}
The details of the Bethe-Salpeter equation and its solution are presented in Appendix \ref{app:BS}. As the Bethe-Salpeter equation is well-known, we simply present the results here. At $T=0$, the general form of $G_2(\q,i\nu)$ is

\begin{equation}\label{collmodeprop}
\big[G_2(i\nu,\q)\big]^{\k,\alpha\beta}_{\k',\lambda\sigma} = \sum_s \varphi^s_{\q,\alpha\beta}(\k)\frac{1}{\omega_{\q,s} - i\nu \,\eta_{\q,s}}\varphi^{s*}_{\q,\lambda\sigma}(\k')\, ,
\end{equation}
where $\omega_{\q,s}$ are the energies of the RPA collective modes, which are guaranteed to be real and positive if the Slater determinant in Eq. \eqref{SlaterDet} is a local energy minimum, i.e. if it satisfies the Hartree-Fock self-consistency equations \eqref{selfconst}. The sum in Eq. \eqref{collmodeprop} runs only over those modes which have non-zero energy. Eq. \eqref{collmodeprop} shows that $G_2(\q,i\nu)$ takes the form of a propagator for the collective modes, which consist of electron-hole pairs with a pair-wavefunction $\varphi^s_{\q,\alpha\beta}(\k)$. 

The collective mode wavefunctions $\varphi^s_{\q,\alpha\beta}(\k)$ and energies $\omega_{\q,s}$ are obtained by solving the following generalized eigenvalue equation:

\begin{equation}\label{collmode}
\begin{split}
    & (n_\alpha(\k) -  n_\beta(\k-\q))\varphi^s_{\q,\alpha\beta}(\k)\eta_{\q,s}\omega_{\q,s}  = \\
    & |E_\alpha(\k)- E_\beta(\k-\q)|\varphi^s_{\q,\alpha\beta}(\k) \\ 
    & + V(\q)\frac{1}{N}\sum_{\k'}\text{tr}\left(\varphi^s_\q(\k')\Lambda_{\q}(\k') \right)\left[ \Lambda^\dagger_{\q}(\k)\right]_{\alpha\beta} \\
    & - \frac{1}{N}\sum_{\q'}V(\q')  \left[\Lambda^\dagger_{\q'}(\k)\varphi^s_{\q}(\k-\q')\Lambda_{\q'}(\k-\q) \right]_{\alpha\beta}\,.
\end{split}
\end{equation}
Note that $\varphi^s_{\q,\alpha\beta}(\k)$ is defined to be zero when $n_\alpha(\k)= n_{\beta}(\k-\q)$. For collective modes with non-zero energy $\omega_{\q,s}$, the signs $\eta_{\q,s}$ are related to the collective mode wavefunctions through the following relation:

\begin{equation}\label{symplnorm}
\sum_\k \sum_{i,n} \left(|\varphi^s_{\q,ni}(\k)|^2 - |\varphi^s_{\q,in}(\k)|^2 \right) = \eta_{\q,s} = \pm 1\,,
\end{equation}
where we have again used the convention that $i$ labels occupied states, and $n$ labels empty states. The physical meaning of the $\eta_{\q,s}$ sign factors and their appearance in the denominator of the collective mode propagator in Eq. \eqref{collmodeprop} can be understood as follows. The Bethe-Salpeter equation at momentum $\q$ describes both the collective mode creation operators, which we denote as $b_{\q}^\dagger$, and the collective mode annihilation operators, denoted as $b_{-\q}$. The collective mode propagator therefore contains contributions of the form $\int\mathrm{d}\tau e^{i\nu\tau} \langle \hat{T}b_{-\k}(\tau)b^\dagger_{-\k}(0)\rangle$ and $\int\mathrm{d}\tau e^{i\nu\tau} \langle \hat{T}b^\dagger_{\k}(\tau)b_{\k}(0)\rangle$, where $\hat{T}$ is the time-ordering operator. The latter can be rewritten as $\int\mathrm{d}\tau e^{-i\nu\tau} \langle \hat{T}b_{\k}(\tau)b^\dagger_{\k}(0)\rangle$, hence the origin of the minus sign in front of $i\nu$ in some of the collective mode propagators. From this we conclude that $\eta_{\q,s} = 1$ means that $\varphi^s_\q$ corresponds to a collective mode creation operator, whereas $\eta_{\q,s}=-1$ means that $\varphi^s_\q$ corresponds to a collective mode annihilation operator.

Now that we have completely defined the collective mode propagator in Eq. \eqref{collmodeprop} as a solution to the Bethe-Salpeter equation it is straightforward to obtain the RPA effective interaction by plugging this propagator in the diagrams in Fig. \ref{fig:Veff} (a). The last four diagrams in this figure can then be interpreted as the effective interaction between electrons which results from the exchange of collective modes. In the next section we will do this explicitly for the collective modes that correspond to Goldstone modes of the spontaneously broken continuous symmetry.

Before ending this section, we mention two facts that will be important for our calculations in the sections below (more details can be found in the appendices):

\begin{itemize}
\item The collective mode propagator is guaranteed to have a particle-hole symmetry, which implies that if $\varphi^s_{\q}$ is a solution to the generalized eigenvalue equation \eqref{collmode} with energy $\omega_{\q,s}$, then so is its particle-hole conjugate partner $\varphi^{s'}_{-\q}$ defined as
\begin{equation}\label{PH}
\varphi^{s'}_{-\q,\alpha\beta}(\k) := \mathcal{P}\left[ \varphi^s_{\q,\alpha\beta}(\k)\right] := \varphi^{s*}_{\q,\beta\alpha}(\k+\q) \,, 
\end{equation}
where $\mathcal{P}$ is the (anti-unitary) particle-hole conjugation operator. Note that the particle-hole transformation inverts the momentum $\q$, but preserves the energy $\omega_{\q,s}$. From Eq. \eqref{symplnorm} we also see that the particle-hole transformation flips the signs $\eta_{\q,s}$. 

\item When the self-consistent Hartree-Fock state has a time-reversal symmetry $\mathcal{T}$, the combined action $\mathcal{PT}$ is a unitary symmetry which preserves momentum but flips the signs $\eta_{\q,s}$ defined in Eq. \eqref{symplnorm}. The $\mathcal{PT}$ symmetry guarantees that non-zero energies appear in degenerate pairs $\omega_{\q,s} = \omega_{\q,s'}$. Furthermore, the corresponding wavefunctions are related by
\begin{equation}\label{PT}
    \varphi^{s'}_{\q,\alpha\beta}(\k) := \mathcal{PT}[\varphi^s_{\q,\alpha\beta}(\k)] := \varphi^s_{\q,\beta\alpha}(-\k+\q)\,,
\end{equation}
where we have without loss of generality chosen to work in a gauge where $\mathcal{T}|u_\alpha(\k)\rangle = |u_\alpha(-\k)\rangle$. Below, we will exclusively focus on time-reversal symmetric self-consistent Hartree-Fock states and make use of this property. In particular, we will use that for every non-zero collective mode energy $\omega_{\q,s}$, there are two collective mode wavefunctions $\varphi^s_{\q}$ which are related by the action of $\mathcal{PT}$, and which have opposite signs $\eta_{\q,s}$. We will use the convention that for every $s>0$, the wavefunctions corresponding to $\omega_{\q,s}$ are denoted as $\varphi^{\pm s}_\q$, where the sign is determined by the corresponding value of $\eta_{\q,s}$. 
\end{itemize}


\section{Interacting electrons and Goldstone modes}\label{sec:interacting}

If the self-consistent Hartree-Fock state spontaneously breaks a continuous symmetry the collective mode spectrum is guaranteed to have gapless branches corresponding to the Goldstone modes. From now on we will exclusively focus on these Goldstone modes, and ignore the other collective modes. We are mainly interested in linearly dispersing Goldstone modes (quadratically dispersing Goldstone modes are simpler, and do not require the full power of the formalism that we develop here). So we assume that at small momenta the Goldstone mode energies can be written as $\omega_{\q,s} = c_s |\q| + \mathcal{O}(\q^2)$, where $c_s$ are the Goldstone mode velocities. 

For linearly dispersing Goldstone modes, there is a one-to-one correspondence between the Goldstone modes and the broken symmetry generators $Q_s$ \cite{Watanabe2012}. For each Goldstone mode we can write down the exact analytic expression for the zero mode wavefunction at $\q = 0$:
\begin{equation}\label{zeromode}
    \tilde{Q}_{s,\alpha\beta}(\k) = i\langle u_\alpha(\k)|Q_s|u_\beta(\k)\rangle \left[n_\alpha(\k)-n_\beta(\k)\right]\,,
\end{equation}
where the factor $i$ is added to ensure that $\tilde{Q}_{s,\alpha\beta}(\k)$ is an eigenstate of the particle-hole conjugation operator $\mathcal{P}$ defined in Eq. \eqref{PH} with eigenvalue $1$. Given that $Q_s$ commutes with $h(\k)$, one can check that $\tilde{Q}_s$ is indeed a zero energy solution of the generalized eigenvalue equation for the collective modes [Eq. \eqref{collmode}]. Furthermore, due to the factor $\left[n_\alpha(\k)-n_\beta(\k)\right]$ the wavefunction $\tilde{Q}_{s,\alpha\beta}(\k)$ is non-zero only when the symmetry generator $Q_s$ is broken (if it is not broken, then $Q_s$ commutes with the mean-field Hamiltonian and hence is diagonal in the mean-field basis $|u_\alpha(\k)\rangle$).

In the remainder of this section we first derive the scattering vertex between electrons and Goldstone modes, and then construct an effective electron-boson model.

\subsection{Electron-Goldstone scattering vertex}


To obtain the interaction between electrons mediated by the exchange of Goldstone modes we start from Eq. \eqref{collmodeprop}, which gives the general form of $G_2(\q,i\nu)$ as a solution to the Bethe-Salpeter equation, and only keep those collective modes that correspond to the Goldstone modes. Taking this expression, and plugging it into the diagrams in Fig. \ref{fig:Veff} (a), we see that the effective RPA interaction is given by the bare interaction (the left most diagram in Fig. \ref{fig:Veff} (a)), plus an interaction where the electrons emit and absorb a Goldstone mode, which can be denoted by the following single diagram:

\begin{equation}\label{VGS}
    \includegraphics[scale=0.9]{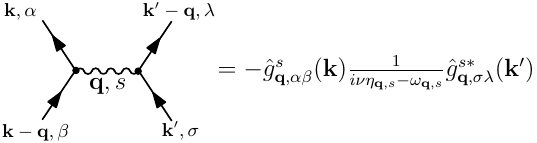}
\end{equation}
In this diagram, the wavy line denotes the boson progagator $(i\nu\eta_{\q,s}-\omega_{\q,s})^{-1}$, and the scattering vertex is given by
\begin{equation}\label{vertex}
    \includegraphics[scale=0.9]{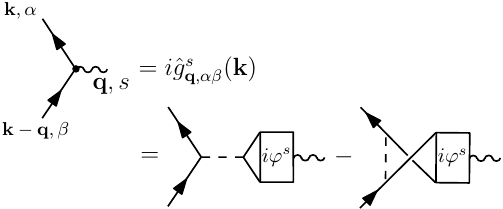}\,,
\end{equation}
where the box represents the Goldstone wavefunction $\varphi^s_{\q,\alpha\beta}(\k)$. Written out in equations, the vertex is given by
\begin{equation}\label{ghat}
\begin{split}
    \hat{g}^s_{\q,\alpha\beta}(\k) = V(\q)\frac{1}{N}\sum_{\k'}&\text{tr}\left(\varphi^s_\q(\k')\Lambda_{\q}(\k') \right)\left[ \Lambda^\dagger_{\q}(\k)\right]_{\alpha\beta} \\
    - \frac{1}{N}\sum_{\q'}V(\q') & \left[\Lambda^\dagger_{\q'}(\k)\varphi^s_{\q}(\k-\q')\Lambda_{\q'}(\k-\q) \right]_{\alpha\beta}
\end{split}
\end{equation}
Note that $\hat{g}^s_{\q,\alpha\beta}(\k)$ is generically non-zero for all $\alpha$ and $\beta$, whereas $\varphi^s_{\q,\alpha\beta}(\k)$ is only non-zero if $n_\alpha(\k)\neq n_\beta(\k-\q)$. The terms in Eq. \eqref{ghat} which define $\hat{g}_\q^s$ have appeared previously as part of the generalized eigenvalue equation in Eq. \eqref{collmode}. Since the Goldstone wavefunctions $\varphi_{\q}^s$ are obtained as solutions to that equation, we immediately obtain from Eqs. \eqref{ghat} and \eqref{collmode} that

\begin{equation}\label{ghat2}
\begin{split}
    \hat{g}^s_{\q,\alpha\beta}(\k) & = \bigg([n_\alpha(\k)-n_\beta(\k-\q)]\eta_{\q,s}\omega_{\q,s}  \\
    & -  |E_\alpha(\k)-E_\beta(\k-\q)|\bigg)\varphi^s_{\q,\alpha\beta}(\k) \\
    \text{ if } & n_\alpha(\k)\neq n_\beta(\k-\q)\, .
\end{split}
\end{equation}
For the components $\hat{g}^s_{\q,\alpha\beta}(\k)$ with $n_\alpha(\k)=n_\beta(\k-\q)$ we have no general analytic expression. 

As explained in Appendix \ref{sec:properties}, we fix the gauge of the Goldstone mode wavefunctions such that the electron-Goldstone vertex satisfies

\begin{equation}\label{Pghat}
    \hat{g}^{s*}_{\q,\alpha\beta}(\k) = \hat{g}^{-s}_{-\q,\beta\alpha}(\k-\q)\,,
\end{equation}
which is a direct consequence of the collective-mode particle-hole symmetry $\mathcal{P}.$ This property of the vertex ensures that the Goldstone-mediated interaction between electrons is Hermitian.

\subsection{Effective electron-boson model}

In the previous section we have derived an explicit expression for the interaction between electrons mediated by the exchange of Goldstone modes. The Goldstone modes are collective excitations of the electron fluid, and hence are not described by independent microscopic degrees of freedom. In this section, however, we will write down an effective theory which does contain both independent electronic and bosonic degrees of freedom, and which mimics the behaviour of the coupled electron-Goldstone-boson system. The action for the effective electron-boson model is given by

\begin{equation}\label{efftheory}
    S = S_{el} + S_{V} + S_B + S_{el-B} + S_C\,,
\end{equation}
where $S_{el}$ and $S_{V}$ respectively denote the quadratic and interacting part of the electronic action that appeared previously in the Hartree-Fock path integral in Eq. \eqref{action}.

The terms $S_B$, $S_{el-B}$, and $S_C$ are not present in the Hartree-Fock action. Of these, the first two describe the Goldstone dynamics and the electron-Goldstone-mode scattering. The Goldstone dynamics is described by

\begin{equation}
    S_B = \int_0^{1/T}\mathrm{d}\tau\, \sum_{s>0}\sum_\q \bar{b}_{\q,s}(\partial_\tau + \omega_{\q,s})b_{\q,s}\,,
\end{equation}
where $\bar{b}_{\q,s},b_{\q,s}$ are bosonic fields corresponding to the Goldstone creation and annihilation operators. The electron-Goldstone coupling is given by

\begin{equation}
\begin{split}
    S_{el-B} & =  \int_0^{1/T}\mathrm{d}\tau\, \frac{1}{\sqrt{N}}\sum_{\q,\k}\bar{\psi}_{\k,\alpha}\psi_{\k-\q,\beta} \times\\
    \sum_{s>0}&\sqrt{\frac{c_s}{2\omega_\q aw_s}}\left( \tilde{g}^s_{\q,\alpha\beta}(\k)\bar{b}_{-\q,s} + \tilde{g}^{-s}_{\q,\alpha\beta}(\k) b_{\q,s}\right)\,,
\end{split}
\end{equation}
where for future use we have defined rescaled vertices:

\begin{equation}\label{rescaledg}
\tilde{g}^{\pm s}_{\q,\alpha\beta}(\k) := \hat{g}^{\pm s}_{\q,\alpha\beta}(\k)\sqrt{\frac{2\omega_\q aw_s N}{c_s}}\,,
\end{equation}
where $a$ is the lattice constant and $w_s$ a dimensionless number. Below we will explain how to choose $w_s$. From Eq. \eqref{Pghat} it follows that tree-level boson exchange in the effective electron-boson model indeed generates the interaction in Eq. \eqref{VGS}.

So far we have taken the Hartree-Fock action, and supplemented it with the bosonic degrees of freedom. The theories with and without the bosonic fields are obviously not the same. In particular, the bosonic degrees of freedom are designed to capture the RPA fluctuations which result from the electronic interaction. As a result, working at tree level with the effective action in Eq. \eqref{efftheory} (without $S_C$) is the same as doing the infinite sum of RPA diagrams in the original theory. Once loops are taken into account in the effective theory, however, one has to be careful that the loop diagrams do not correspond to a double counting of RPA diagrams in the original theory. For example, $\omega_{\q,s}$ is already the complete boson dispersion which is generated by the electron dynamics, and it should therefore not be further renormalized by coupling to the electrons. To ensure this, we have to add a counter term which is quadratic in the boson fields, and which removes all self-energy diagrams for the bosons. Another counter term quartic in the fermion fields has to be added to remove the RPA renormalization of the electron repulsion, as this effect is already contained in the interaction mediated by boson exchange. All these counter terms are contained in $S_C$. It is not necessary to construct these terms explicitly -- all one needs to know is that they eliminate certain diagrams such that working with the effective theory beyond tree level is the same as working with the original theory.

\subsection{Transforming to the real basis}

In this final section we bring the effective electron-boson model introduced in the previous section in a more standard form. First, we rescale the boson fields as follows
\begin{equation}
    \bar{b}_{\q,s},b_{\q,s}\rightarrow \sqrt{\frac{aw_s}{c_s}}\bar{b}_{\q,s}, \sqrt{\frac{aw_s}{c_s}}b_{\q,s}\,,
\end{equation}
and then define the usual (Fourier transformed) canonically conjugate real fields as
\begin{eqnarray}
    \phi_{\q,s} & = & \frac{1}{\sqrt{2\omega_\q}}(\bar{b}_{\q,s} + b_{-\q,s}) \\
    \pi_{\q,s} & = & i\sqrt{\frac{\omega_\q}{2}}(\bar{b}_{\q,s} - b_{-\q,s})\,,
\end{eqnarray}
where we adopted the standard notation. In terms of the new fields, the quadratic boson action becomes
\begin{equation}
\begin{split}
S_B = \int_0^{1/T}\mathrm{d}\tau & \sum_{s>0}\frac{aw_s}{c_s} \sum_\q\bigg(-i\pi_{\q,s}\partial_\tau \phi_{-\q,s} \\
&+ \frac{1}{2}\pi_{\q,s}\pi_{-\q,s} + \frac{1}{2}\omega_{\q,s}^2 \phi_{\q,s}\phi_{-\q,s} \bigg)\,,
\end{split}
\end{equation}
and the electron-boson vertex is given by
\begin{equation}
\begin{split}
    S_{el-B} =&  \int_0^{1/T}\mathrm{d}\tau\, \frac{1}{\sqrt{N}}\sum_{\q,\k}\bar{\psi}_{\k,\alpha}\psi_{\k-\q,\beta} \times\\
    \sum_{s>0}&\left( g^s_{\q,\alpha\beta}(\k)\phi_{-\q,s} + f^{s}_{\q,\alpha\beta}(\k) \pi_{-\q,s}\right)\,,
\end{split}
\end{equation}
with
\begin{eqnarray}
    g^s_{\q,\alpha\beta}(\k) & = & \frac{1}{2}\left(\tilde{g}^s_{\q,\alpha\beta}(\k) + \tilde{g}^{-s}_{\q,\alpha\beta}(\k)\right)\label{gs} \\
    f^s_{\q,\alpha\beta}(\k) & = & \frac{-i}{2\omega_{\q,s}}\left(\tilde{g}^s_{\q,\alpha\beta}(\k) - \tilde{g}^{-s}_{\q,\alpha\beta}(\k)\right)\,.\label{fs}
\end{eqnarray}
As a final step, we integrate out the $\pi_{\q,s}$ fields. This produces the following quadratic boson action:
\begin{equation}
    S_B = \frac{1}{2}\int\mathrm{d}\tau\,\sum_{s>0}\frac{aw_s}{c_s}\sum_\q\left( -\phi_{-\q}\partial^2_\tau \phi_\q + \omega_{\q,s}^2 \phi_{\q,s}\phi_{-\q,s}\right)\, .
\end{equation}
In the long-wavelength continuum limit, using $\omega_{\q,s} \sim c_s^2\q^2$, we can write this as the standard action for relativistic real scalar fields:
\begin{equation}
S_B = \frac{1}{2}\int\mathrm{d}\tau\int \mathrm{d}^2\r\, \sum_{s>0} \chi_s (\partial_\tau \phi_s)^2 + \rho_s (\nabla \phi_s)^2\,,
\end{equation}
with $\chi_s = w_s/ac_s$, and $\rho_s = w_s c_s/a$ is the stiffness. After integrating out $\pi_{\q,s}$, the electron-Goldstone vertex becomes
\begin{equation}
\begin{split}
    S_{el-B} =&  \int_0^{1/T}\mathrm{d}\tau\, \frac{1}{\sqrt{N}}\sum_{\q,\k}\bar{\psi}_{\k,\alpha}\psi_{\k-\q,\beta} \times\\
    \sum_{s>0}&\left( g^s_{\q,\alpha\beta}(\k)\phi_{-\q,s} + f^{s}_{\q,\alpha\beta}(\k) i\partial_\tau \phi_{-\q,s}\right)\,,
\end{split}
\end{equation}
where in the second term $\pi_{\q,s}$ is now replaced with $i\partial_\tau \phi_{\q,s}$. Finally, we also obtain following two-body interaction from integrating out $\pi_{\q,s}$:
\begin{equation}\label{Spi}
\begin{split}
    S_{V\pi} =\int\mathrm{d}\tau  \sum_{s>0}\frac{-c_s}{2aw_s N}&\sum_{\k,\k',\q}\big(\bar{\psi}_\k f^s_\q(\k)\psi_{\k-\q} \big)\times \\
    & \left(\bar{\psi}_{\k'} f^s_{-\q}(\k')\psi_{\k'+\q} \right)\,,
\end{split}
\end{equation}
where we have suppressed the Greek indices, and instead adopted a matrix notation. From Eqs. \eqref{fs} and \eqref{Pghat} it follows that we can equivalently write Eq. \eqref{Spi} as
\begin{equation}
\begin{split}\label{SVpi}
    S_{V\pi} = \int\mathrm{d}\tau \sum_{s>0} \frac{-c_s}{2aw_s N}&\sum_{\k,\k',\q}\big(\bar{\psi}_\k f^s_\q(\k)\psi_{\k-\q} \big)\times \\
    & \left(\bar{\psi}_{\k'-\q} f_{\q}^{s\dagger}(\k')\psi_{\k'} \right)\,,
\end{split}
\end{equation}
which makes explicit that this is a negative definite interaction.

\section{The rotating frame}\label{sec:rotframe}

We continue to work with the effective electron-boson model derived in the previous section. Even though this model was shown to exactly reproduce the RPA result, we will argue below that it has an (apparent) unphysical property, which can be removed by working in the so-called `rotating frame'.

\subsection{The problem: strong interband scattering in the $\q\rightarrow 0$ limit}

It follows from Eqs. \eqref{ghat2} and \eqref{rescaledg} that the vertices $g^s_\q$ and $f^s_{\q}$ in the effective electron-boson model [defined in Eqs. \eqref{gs} and \eqref{fs}] are given by
\begin{widetext}
\begin{eqnarray}\label{explicit}
    g^s_{\q,\alpha\beta}(\k) & = & \frac{1}{2}\bigg( [n_\alpha(\k)-n_\beta(\k-\q)]\omega_{\q,s}\left( \tilde{\varphi}^s_{\q,\alpha\beta}(\k)- \tilde{\varphi}^{-s}_{\q,\alpha\beta}(\k)\right)
    -|E_\alpha(\k)-E_\beta(\k-\q)|\left(  \tilde{\varphi}^s_{\q,\alpha\beta}(\k)+\tilde{\varphi}^{-s}_{\q,\alpha\beta}(\k)\right)\bigg) \\
    f^s_{\q,\alpha\beta}(\k) & = & \frac{-i}{2\omega_{\q,s}}\bigg( [n_\alpha(\k)-n_\beta(\k-\q)]\omega_{\q,s}\left( \tilde{\varphi}^s_{\q,\alpha\beta}(\k)+\tilde{\varphi}^{-s}_{\q,\alpha\beta}(\k)\right)
    -|E_\alpha(\k)-E_\beta(\k-\q)|\left(  \tilde{\varphi}^s_{\q,\alpha\beta}(\k)-\tilde{\varphi}^{-s}_{\q,\alpha\beta}(\k)\right)\bigg)\nonumber\,,
\end{eqnarray}
if $n_\alpha(\k)\neq n_\beta(\k-\q)$. Here we have used that $\eta_{\q,s}=-\eta_{\q,-s}$, and we have defined
\end{widetext}
\begin{equation}\label{rescaled}
\tilde{\varphi}^{\pm s}_{\q,\alpha\beta}(\k) := \varphi^{\pm s}_{\q,\alpha\beta}(\k)\sqrt{\frac{2\omega_\q aw_s N}{c_s}}\,.
\end{equation}
In Appendix \ref{sec:properties} we show that in a suitable gauge for the collective mode wavefunctions one can choose $w_s$ such that the following equation holds:
\begin{equation}\label{lim}
\lim_{\q\rightarrow 0} \tilde{\varphi}^{\pm s}_{\q,\alpha\beta}(\k) = \tilde{Q}_{s,\alpha\beta}(\k)\,,
\end{equation}
where $\tilde{Q}_{s,\alpha\beta}(\k)$ is the Goldstone zero mode at $\q = 0$ defined in Eq. \eqref{zeromode}. From the exact expression for this zero mode, and Eq. \eqref{lim}, we see that 
\begin{equation}\label{gslim}
    \lim_{\q\rightarrow 0} g^s_{\q,\alpha\beta}(\k) = i[E_\alpha(\k)-E_\beta(\k)]\langle u_\alpha(\k)|Q_s|u_\beta(\k)\rangle\,,
\end{equation}
if $n_\alpha(\k)\neq n_\beta(\k)$. Equation \eqref{gslim} shows that $\q = 0$ Goldstone modes induce scattering between electrons in different mean-field bands. This is because the mean-field bands have gaps which result from the presence of a non-zero symmetry-breaking order parameter. The way the original, non-interacting bands are split to generate the mean-field bands depends on the orientation of the order parameter. So a $\q=0$ Goldstone mode, which corresponds to a global rotation of the order parameter, will induce a mixing of the mean-field bands which enter the Hartree-Fock path integral, as they are defined using an order parameter with a chosen fixed orientation. We note that Eq. \eqref{gslim} has previously also been obtained in Ref. \cite{Watanabe2014}, although the authors of that work did not use the RPA approach adopted here.

The exchange of long-wavelength Goldstone modes generates an interaction between mean-field electrons in different bands which is of the form $\sim \Delta^2/(\chi_s (i\nu)^2 - \rho_s \q^2)$, where $\Delta$ is the gap between the mean-field bands induced by the symmetry breaking. As a result, the energy scale associated with a non-resonant process where the exchange of a long wavelength Goldstone mode with energy $c_s|\q| = \sqrt{\rho_s/\chi_s} |\q| \ll \Delta$ scatters a pair of electrons across the bandgap is $~\Delta^2/(\chi_s \Delta^2 -\rho_s\q^2) \sim \chi_s^{-1}$. As $\chi_s^{-1}$ is generically expected to be order one in natural energy scales (which is indeed the case in the examples discussed below), such non-resonant processes cannot be ignored. The inevitable conclusion is thus that generically the mean-field electrons are strongly coupled to the long wavelength Goldstone modes. However, in the next section we will show that electrons which are defined in a rotating frame are not affected by small-$\q$ Goldstone modes. 

As a final comment, note that from Eq. \eqref{explicit} it follows that also $\lim_{\q\rightarrow 0}f^s_\q \neq 0$. However, as $f^s_\q$ is the vertex which couples the mean-field electrons to $\partial_\tau \phi$, this leads to an interaction of the form $\sim \nu^2/((i\nu)^2 - c^2\q^2)$, which remains finite in the $\nu,\q\rightarrow 0$ limit.

\subsection{The solution: the rotating frame}

To remove the strong inter-band interactions between the mean-field electrons discussed in the previous section we perform a change of integration variables in the path integral of the electron-boson model. In particular, in the local orbital basis we change the Grassmann fields as
\begin{equation}\label{realspace}
\psi_a(\r) \rightarrow \sum_b R_{ab}(\r)\psi_b(\r)\,,
\end{equation}
with
\begin{equation}
    R(\r) = \exp\left(-i\sum_{s>0}\phi_s(\r)Q_s \right)\, .
\end{equation}
As this is a unitary transformation, the corresponding change of integration variables has a trivial Jacobian. In momentum space, and working in the mean-field basis, the transformation in Eq. \eqref{realspace} can be written as

\begin{equation}\label{linearord}
    \psi_\alpha(\k) \rightarrow \psi_\alpha(\k) - \frac{i}{\sqrt{N}}\sum_{s,\q,\beta} \phi_{\q,s} Q^s_{\q,\alpha\beta}(\k)\psi_\beta(\k-\q) + \mathcal{O}(\phi^2)\,,
\end{equation}
where we have defined
\begin{equation}
   Q^s_{\q,\alpha\beta}(\k) = \langle u_\alpha(\k)|Q_s|u_\beta(\k-\q)\rangle\, .
\end{equation}
Performing this change of variables in the kinetic term of the mean-field fermions, $\sum_{\k,\alpha} \bar{\psi}_{\k,\alpha}(\partial_\tau + E_\alpha(\k))\psi_{\k,\alpha}$, produces terms which can be absorbed in $S_{el-B}$. In particular, using Eq. \eqref{linearord} we find that the change of Grassmann variables induces following shift in the vertex functions:

\begin{eqnarray}
    g^s_{\q,\alpha\beta}(\k) & \rightarrow &g^s_{\q,\alpha\beta}(\k) - iQ^s_{\q,\alpha\beta}(\k)[E_\alpha(\k)-E_\beta(\k-\q)] \nonumber \\
    & & =: g^{Rs}_{\q,\alpha\beta}(\k) \label{gR} \\
    f^s_{\q,\alpha\beta}(\k) & \rightarrow & f^s_{\q,\alpha\beta}(\k) -Q^s_{\q,\alpha\beta}(\k) \nonumber \\
    && =: f^{Rs}_{\q,\alpha\beta}(\k) \label{fR}
\end{eqnarray}
The change of variables also introduces higher-order interaction terms of the form $\phi^n \bar{\psi}\psi$ with $n>1$, but we ignore these terms here.

From Eqs. \eqref{gslim} and \eqref{gR} we see that 

\begin{equation}
    \lim_{\q\rightarrow 0} g^{Rs}_{\q,\alpha\beta}(\k) = 0\,,
\end{equation}
i.e. the $\nu,\q = 0$ Goldstone modes decouple from the electrons. This shows that the electrons defined in Eq. \eqref{realspace}, which live in a rotating frame that locally follows the order parameter fluctuations, will have much smaller self-energy corrections than the mean-field electrons, and hence will be much closer to the true quasi-particles of the symmetry-broken system.

Before concluding this section let us mention the effect of going to the rotating frame on the other terms in the action of the electron-boson model. First, the change of Grassmann variables in Eq. \eqref{realspace} leaves the bare interaction term $S_V$ invariant. However, in general it will change the interaction contained in $S_{V\pi}$ \eqref{SVpi}. For some applications it might be important to keep this in mind, but in the remainder of this work we will not use the interactions contained in $S_V$ and $S_{V\pi}$ anymore.

\subsection{Implications for electron spectral functions}

In this section we investigate the consequences of the fact that quasi-particles are defined in a rotating frame for the correlation functions of microscopic fermions. In particular, we are interested in the two-point functions 
\begin{eqnarray}
    G_{ab}(\tau,\r-\r') & = & -\langle \hat{T} c_{\r,a}(\tau)c^\dagger_{\r',b}(0)\rangle_\beta \\
    & = & \frac{-1}{Z}\int_{\bar{\psi},\psi,\phi} \psi_{\r,a}\bar{\psi}_{\r',b} \,e^{-S[\bar{\psi},\psi,\phi]}\,,\label{Sunrotated}
\end{eqnarray}
where $\hat{T}$ is again the time-ordering operator, $\langle \cdot \rangle_\beta$ is the thermal average, and $\int_{\bar{\psi},\psi,\phi}$ a path integral for the fields $\bar{\psi},\psi$ and $\phi$. Note that the fermion fields in Eq. \eqref{Sunrotated} are in the original, unrotated frame. Performing the change of integration variables as in Eq. \eqref{realspace}, we obtain 
\begin{equation}
    G_{ab}(\tau,\r) = -\sum_{c,d}\langle R^*_{ca}(\tau,\r) \psi_{c,\r}(\tau) R_{db}(0,0)\bar{\psi}_{d,0}(0) \rangle_\beta \,,\nonumber
\end{equation}
where $\bar{\psi},\psi$ now live in the rotated frame, and hence describe the quasi-particles of the broken-symmetry state. As a first approximation, we ignore the interaction between the electrons and the Goldstone modes, such that the correlation function of the microscopic electrons factorizes as
\begin{equation}
G_{ab}(\tau,\r) = -\sum_{c,d}\langle R^*_{ca}(\tau,\r)R_{db}(0,0)\rangle\, \langle \psi_{c,\r}(\tau) \bar{\psi}_{d}(0,0) \rangle_\beta \,.\nonumber
\end{equation}
By working with this approximation we make a physical assumption that corrections resulting from electron-Goldstone mode interactions will be subleading compared to the above `zeroth order' contribution. This assumption is partly justified by the fact that fermions in the rotating frame decouple from the Goldstone modes at long wavelengths. Going to frequency and momentum space we then obtain 
\begin{equation}\label{conv}
\begin{split}
   G_{ab}(i\omega,\k) = \frac{-T}{N}\sum_{i\nu,\q}&\sum_{cd\alpha} D^R_{ab,cd}(i\nu,\q) \\
   &\times \frac{u^{c}_\alpha(\k-\q)u^{d*}_\alpha(\k-\q)}{i(\omega-\nu)-E_{\k-\q,\alpha}}\,,
   \end{split}
\end{equation}
where $D^R_{ab,cd}(i\nu,\q)=-\langle R^*_{ca}(i\nu,\q)R_{db}(i\nu,\q)\rangle$. We thus find that the fermion correlation function in frequency-momentum space is a convolution of the fermionic quasi-particle propagator with the propagator of the Goldstone modes. Note that Eq. \eqref{conv} has previously been obtained for the half-filled Hubbard model in Ref. \cite{Borejsza2004}, and very similar expressions are found in theories where the electron is assumed to fractionalize in a spinon and a holon \cite{Florens2004,Qi2010,Tang2013,Chatterjee2017,Scheurer2018,Sachdev2019,Bonetti2022}.

Up to now our effective electron-boson model describes a collection of independent generalized `spin-waves', and does not take topological order parameter configurations, such as e.g. vortices or skyrmions, or interactions between Goldstone modes into account. To remedy this, one can improve the model to correctly capture the compact nature of the order parameter manifold by rewriting the Goldstone action as a non-linear sigma model. We will do this explicitly in the next section for the anti-ferromagnetic ground state of the Hubbard model (as was originally done in Refs. \cite{Schulz1990,Schulz1994,Weng}). With the non-linear sigma model description of the Goldstone modes, it is possible that thermal fluctuations destroy the long-range order at a much lower temperature than the mean-field transition temperature at which the symmetry-breaking gap disappears from the Hartree-Fock band spectrum. This is especially relevant in 2D, where the Hohenberg-Mermin-Wagner theorem states that long-range order from continuous symmetry breaking disappears at any non-zero temperature. In the resulting thermally disordered phase, the propagator $D^R_{ab,cd}(i\nu,\q)$ is symmetric, i.e. $D^R_{ab,cd}(i\nu,\q) \propto \delta_{ab}$, and acquires a mass. From Eq. \eqref{conv}, we see that in that case the electron Green's function $G_{ab}(i\omega,\k)$ also becomes symmetric in the thermally disordered phase, even though the order-parameter-induced gap can still be present in the mean-field spectrum of the fermions in the rotating frame. This effect therefore naturally leads to `pseudo-gap' physics in thermally disordered broken-symmetry states. Below we illustrate this explicitly in the two example sections.

\section{Example I: anti-ferromagnetism in the Hubbard model}\label{sec:ex1}

In this first example section we apply the general formalism introduced above to the anti-ferromagnetic ground state of the square-lattice Hubbard model.

\subsection{Mean-field state and Goldstone mode energies}

We are interested in the Hubbard model on the square lattice, which is defined by the following Hamiltonian:

\begin{equation}
    H = -t \sum_{\langle ij\rangle}\sum_s c^\dagger_{i,s} c_{j,s} -t'\sum_{\langle\langle ij\rangle\rangle} c^\dagger_{i,s} c_{j,s} + h.c. + U\sum_i n_{i,\uparrow} n_{i,\downarrow}\,, 
\end{equation}
where the first sum is over nearest neighbors, and the second sum over next nearest neighbors. In the interaction term, $n_{i,s} = c^\dagger_{i,s}c_{i,s}$ is the density of electrons with spin $s$ at site $i$. We choose units of energy such that $t\equiv 1$, and take $t' = -0.35$. 

For our purposes, it suffices to only consider half filling, where it is well-known that the ground state is an insulating anti-ferromagnet for sufficiently large $U$. An anti-ferromagnet (AFM) breaks translation symmetry, so at first sight the general formalism introduced above, which explicitly assumes translation invariance, does not seem to apply. However, an AFM is invariant under the combined action of translating by one lattice site and flipping all the spins. Let's call this modified translation symmetry $T_{x/y}'$. If we consider the AFM state with periodic boundary conditions, i.e. on a torus of size $L_x\times L_y$ (note that both $L_x$ and $L_y$ have to be even), then we see that $T_x^{'L_x} = T_y^{'L_y} = \mathds{1}$. This shows that the eigenvalues of $T'_x$ ($T'_y$) are phases $e^{i\tilde{k}_x}$ ($e^{i\tilde{k}_y})$, with $\tilde{k}_x = \frac{2\pi n}{L_x}$ ($\tilde{k}_y = \frac{2\pi n}{L_y}$), just as for the conventional translation operators. Because the AFM is invariant under $T'_{x/y}$, the correlation functions satisfy $\langle c^\dagger_{\tilde{\k}}c_{\tilde{\k}'}\rangle \propto \delta_{\tilde{\k},\tilde{\k}'}$, i.e. the single-particle density matrix is diagonal in $\tilde{\k}$. So by working in the basis where $T'_{x/y}$ is diagonal, we can treat the AFM in the same way as conventional translationally invariant states, by making use of the conserved pseudo-momentum $\tilde{\k}$.

Let us assume without loss of generality that the AFM order is in the XY plane. $T'_{x/y}$ can then be defined as a translation followed by a $\pi$ rotation along the $z$ axis:

\begin{equation}
    T'_{\r} = e^{i\r\cdot \Q\, s^z/2} T_{\r}\,,
\end{equation}
where $\r = (n,m)$ with $n,m\in\mathbb{Z}$ describes a general lattice vector, $\Q = (\pi,\pi)$, and $s^i$ are the Pauli matrices acting on the spin indices. In this case, the connection between the pseudo-momentum $\tilde{\k}$ and the crystal momentum $\k$ is given by:

\begin{eqnarray}
c^\dagger_{\tilde{\k},\uparrow} & = & c^\dagger_{\k-\Q/2,\uparrow} \\ \label{74}
c^\dagger_{\tilde{\k},\downarrow} & = & c^\dagger_{\k+\Q/2,\downarrow} \label{75}
\end{eqnarray}
The self-consistent Hartree-Fock mean-field Hamiltonian is diagonal in the pseudo-momentum $\tilde{\k}$, and, assuming AFM order along the $x$-direction, can be written as 

\begin{equation}\label{HHF}
H_{HF} = \sum_{\tilde{\k}} c^\dagger_{\tilde{\k}} \left(\begin{matrix} \varepsilon_{\tilde{\k},\uparrow} & \Delta_{\tilde{\k}} \\ \Delta_{\tilde{\k}} & \varepsilon_{\tilde{\k},\downarrow}\end{matrix}\right) c_{\tilde{\k}}\, ,
\end{equation}
where $c^\dagger_{\tilde{\k}} = (c^\dagger_{\tilde{\k},\uparrow}, c^\dagger_{\tilde{\k},\downarrow})$. We have numerically solved the Hartree-Fock self-consistency equations on a $34\times 34$ pseudo-momentum grid using $U = 5$, and we find an AFM order parameter $\Delta_{\tilde{\k}} = \Delta \approx 1.93$.

\begin{figure}
\includegraphics[scale=0.29]{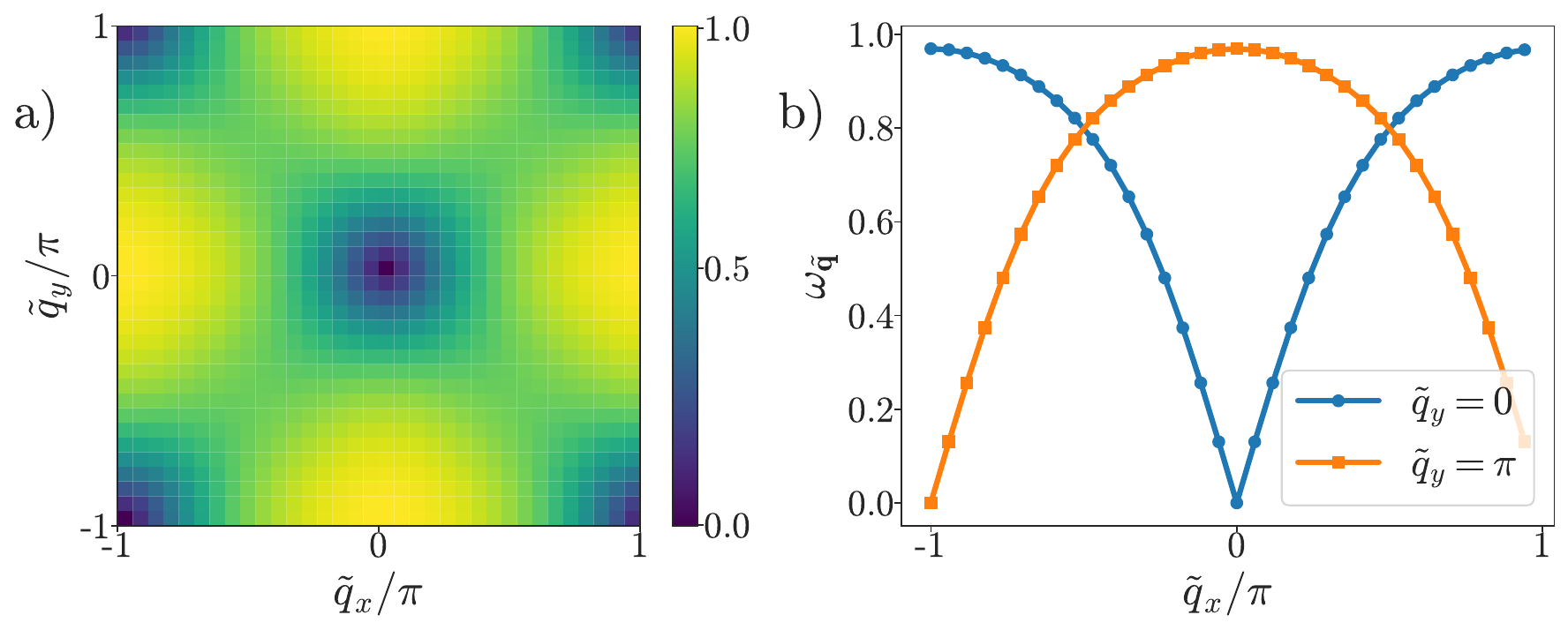}
\caption{Goldstone energy $\omega_{\tilde{\q}}$, obtained by solving the Bethe-Salpeter equation \eqref{collmode} for the square-lattice Hubbard model with $U=5$ on a $34\times 34$ system, as a function of the pseudo-momentum.}\label{fig:omega}
\end{figure}

We next obtain the Goldstone mode energies and wavefunctions by numerically solving the generalized eigenvalue equation in Eq. \eqref{collmode}, where each momentum label $\k$ is to be replaced with a pseudo-momentum label $\tilde{\k}$. The numerically obtained energy for the lowest-energy collective mode (i.e. the Goldstone mode) is shown in Fig. \ref{fig:omega} over the entire pseudo-momentum Brillouin zone, and along two cuts $\tilde{q}_y =0$ and $\tilde{q}_y =\pi$. As expected, we find two linearly dispersing Goldstone modes, one at pseudo-momentum $\tilde{\q}=(0,0)$, and one at $\tilde{\q}=(\pi,\pi)$. As the AFM order is along the $x$-direction, we can understand this by noting that the two broken-symmetry generators $s^z$ and $s^y$ at crystal momentum $\q = (0,0)$ respectively have pseudo-momentum $\tilde{\q}=(0,0)$ and $\tilde{\q}=(\pi,\pi)$. From a fit to the linear part of the dispersion relation $\omega_{\tilde{\q}}$, we find a Goldstone velocity $c \approx 0.988$.

\subsection{Electron-Goldstone scattering vertices}

By solving the generalized eigenvalue equation \eqref{collmode}, we also obtain the collective mode wavefunctions $\varphi^{\pm}_{\tilde{\q},\alpha\beta}(\tilde{\k})$, which can be used to construct the electron-Golstone scattering vertices $g_{\tilde{\q},\alpha\beta}(\tilde{\k})$ and $f_{\tilde{\q},\alpha\beta}(\tilde{\k})$ as explained in Sec. \ref{sec:interacting}. Recall that $\alpha$ and $\beta$ label the mean-field bands, i.e. the eigenstates of $H_{HF}$ defined in Eq. \eqref{HHF}, and not the spin indices. 

To go to the rotating frame, we use

\begin{equation}
R(\r) = \exp\left(-i\left[\phi_z(\r)s^z + \phi_y(\r)(-1)^{\Q\cdot \r} s^y \right] \right)\,,
\end{equation}
where $\phi_z(\r)$ and $\phi_y(\r)$ contain pseudo-momenta $\tilde{\k}$ which lie in the magnetic Brillouin zone, i.e. the Brillouin zone with reciprocal vectors $(\pi,\pm \pi)$. It thus follows that $Q_{\tilde{\q},\alpha\beta}(\tilde{\k})$, as defined in Eq. \eqref{linearord}, is given by

\begin{equation}
Q_{\tilde{\q},\alpha\beta}(\tilde{\k}) = \langle u_\alpha(\tilde{\k})|s^z|u_\beta(\tilde{\k}-\tilde{\q})\rangle\,,
\end{equation}
if $\tilde{\q}$ lies in the first magnetic Brillouin zone, and 

\begin{equation}
Q_{\tilde{\q},\alpha\beta}(\tilde{\k}) = \langle u_\alpha(\tilde{\k})|s^y|u_\beta(\tilde{\k}-\tilde{\q})\rangle\,,
\end{equation}
if $\tilde{\q}$ lies in the complement of the first magnetic Brillouin zone in the full pseudo-momentum Brillouin zone. From Eq. \eqref{HHF}, it is clear that the mean-field states satisfy $|u_\alpha(\tilde{\k}+\Q)\rangle = \pm s^x |u_\alpha(\tilde{\k})\rangle$ (assuming that we work in a gauge where the mean-field states are real). From this it follows that $Q_{\tilde{\q}}$ is periodic under shifts by $\Q$ and multiplication by $i$, up to gauge-dependent factors $\pm 1$. 

The rotating frame is introduced to ensure that the scattering vertex $g^R_{\tilde{\q},\alpha\beta}(\tilde{\k})$, defined in Eq. \eqref{gR}, is zero as $\tilde{\q}\rightarrow 0,\Q$. As explained in Sec. \ref{sec:properties}, this requires both a rescaling of the collective mode wavefunctions by a factor $w^{1/2}$, and a gauge fixing procedure. The gauge fixing procedure is simplified by choosing AFM order along the $x$-direction, such that the collective mode wavefunctions $\varphi_{\tilde{\q}}$ can initially be taken to be real. We then fix the gauge of the $\varphi_{\tilde{\q}}$ by multiplication with $\pm i$ when $\tilde{\q}$ lies in the first magnetic Brillouin zone, and $\pm 1$ otherwise, such that the phase of an arbitrary off-diagonal element in $\varphi_{\tilde{\q},\alpha\beta}(0)$ agrees with the phase of the corresponding off-diagonal element in $iQ_{\tilde{\q},\alpha\beta}(0)[n_\alpha(0) - n_\beta(\tilde{\q})]$. After the gauge fixing, we find that $\lim_{\tilde{\q}\rightarrow 0,\Q}g^R_{\tilde{\q}} = 0$ if we take $w \approx 0.922$.

\begin{figure}
\includegraphics[scale=0.28]{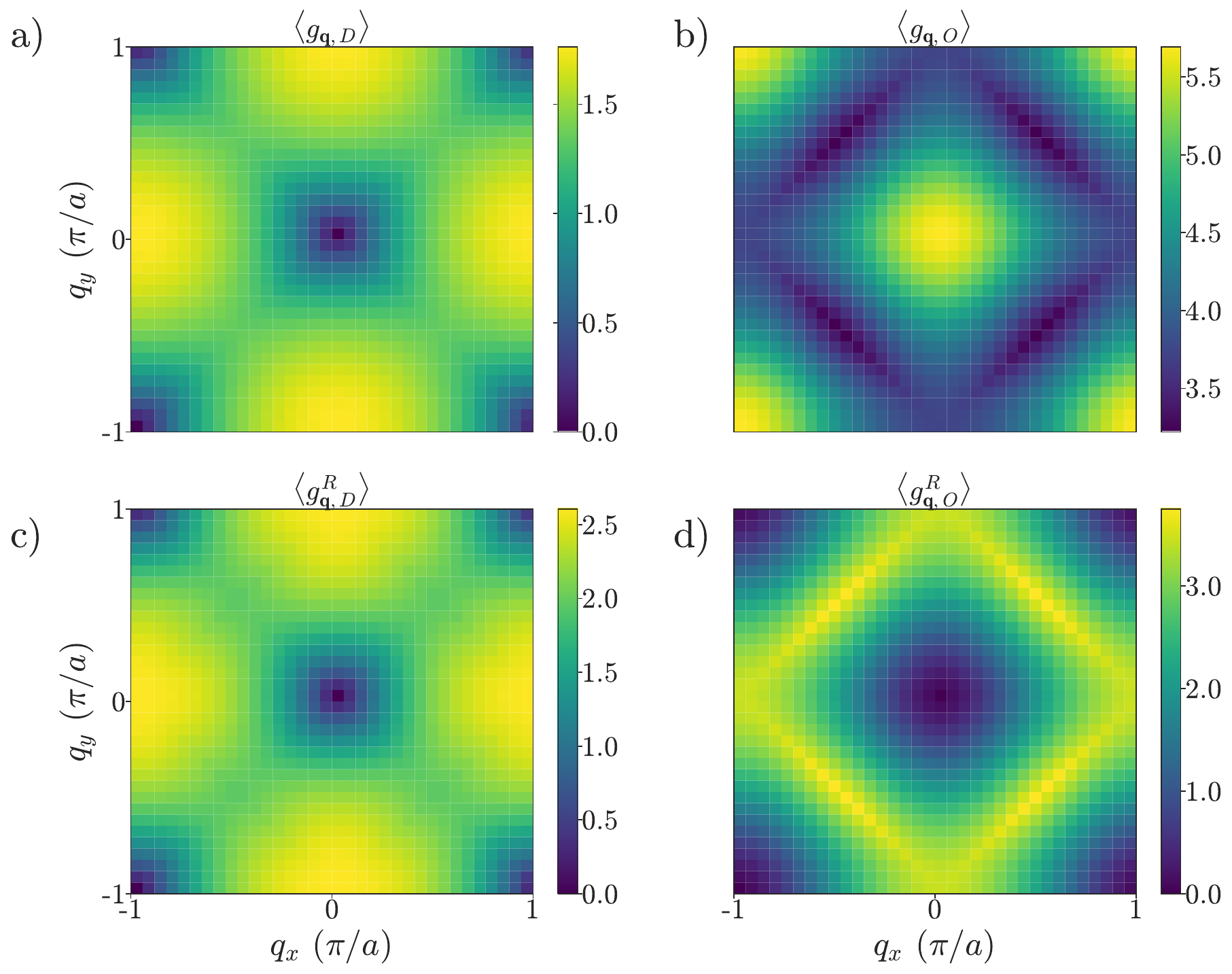}
\caption{Averaged electron-Goldstone scattering vertices $\langle g_{\tilde{\q}}\rangle$. (a) Intra-band vertex in the mean-field basis [Eq. \eqref{gav1}]. (b) Inter-band vertex in the mean-field basis [Eq. \eqref{gav2}]. (c) Intra-band vertex in the rotating frame [Eq. \eqref{gRav1}]. (d) Inter-band vertex in the rotating frame [Eq. \eqref{gRav2}].}\label{fig:gs}
\end{figure}

\begin{figure}
\includegraphics[scale=0.28]{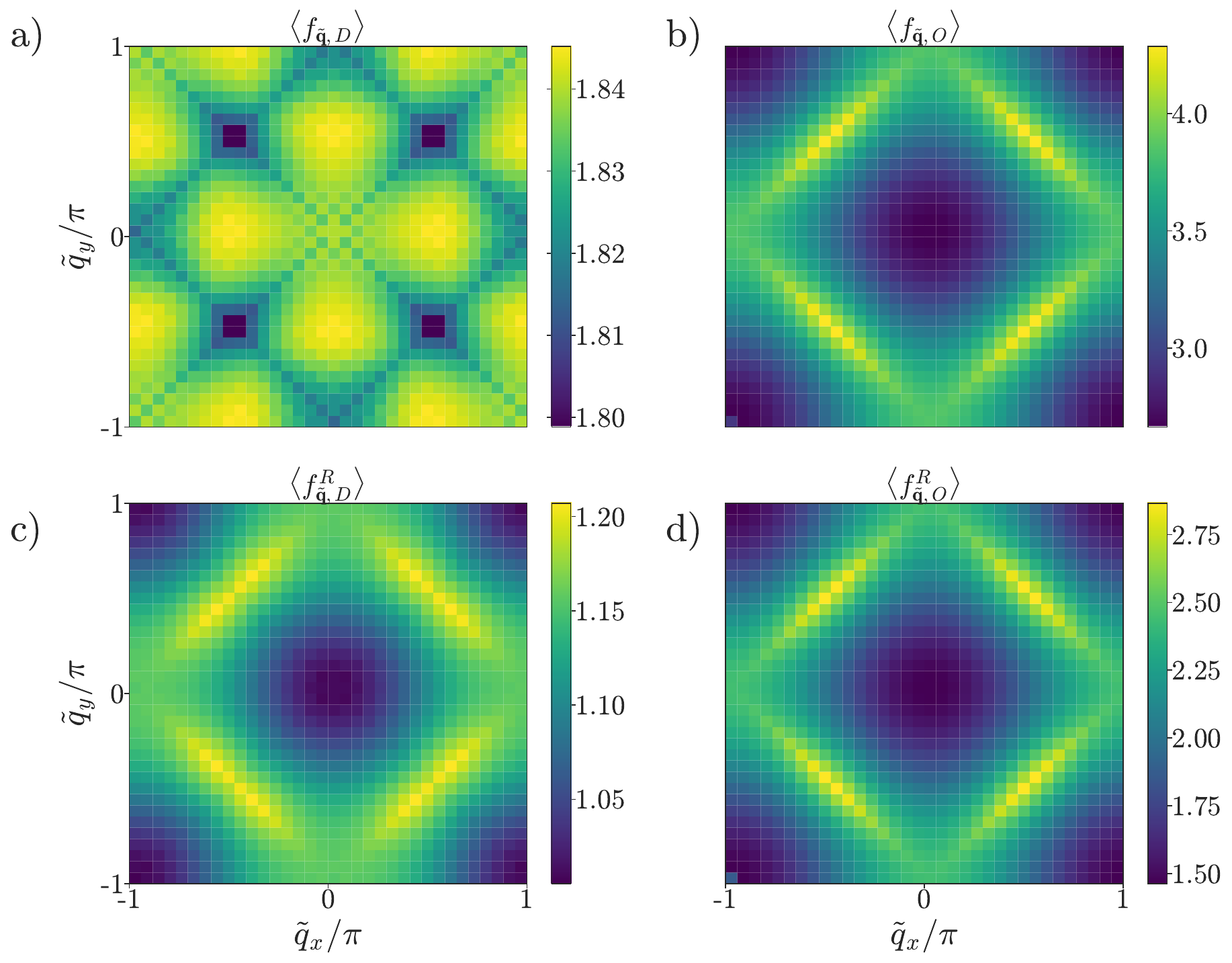}
\caption{Averaged electron-Goldstone scattering vertices $\langle f_{\tilde{\q}}\rangle$. (a) Intra-band vertex in the mean-field basis. (b) Inter-band vertex in the mean-field basis. (c) Intra-band vertex in the rotating frame. (d) Inter-band vertex in the rotating frame.}\label{fig:fs}
\end{figure}

To illustrate our numerical results for the electron-Goldstone scattering vertices, we define

\begin{eqnarray}
\langle g_{\q,D}\rangle^2 & = & \frac{1}{L_xL_y}\sum_{\tilde{\k}} |g_{\tilde{\q},00}(\tilde{\k})|^2 + |g_{\tilde{\q},11}(\tilde{\k})|^2 \, ,\label{gav1}\\
\langle g_{\q,O}\rangle^2 & = & \frac{1}{L_xL_y}\sum_{\tilde{\k}} |g_{\tilde{\q},01}(\tilde{\k})|^2 + |g_{\tilde{\q},10}(\tilde{\k})|^2 \, , \label{gav2}
\end{eqnarray}
as a pseudo-momentum averaged version of respectively the intra-band (diagonal) and inter-band (off-diagonal) scattering vertices in the mean-field basis. Similarly, we define the averaged scattering vertices in the rotating frame as

\begin{eqnarray}
\langle g^R_{\q,D}\rangle^2 & = & \frac{1}{L_xL_y}\sum_{\tilde{\k}} |g^R_{\tilde{\q},00}(\tilde{\k})|^2 + |g^R_{\tilde{\q},11}(\tilde{\k})|^2  \, ,\label{gRav1} \\
\langle g^R_{\q,O}\rangle^2 & = & \frac{1}{L_xL_y}\sum_{\tilde{\k}} |g^R_{\tilde{\q},01}(\tilde{\k})|^2 + |g^R_{\tilde{\q},10}(\tilde{\k})|^2  \, .\label{gRav2}
\end{eqnarray}
In Fig. \ref{fig:gs} (a) and (b) we show the averaged scattering vertices in the mean-field basis. As expected, the intra-band scattering vertex $\langle g_{\tilde{\q},D}\rangle$ goes to zero as $\tilde{\q}\rightarrow 0,\Q$, whereas the inter-band vertex $\langle g_{\tilde{\q},O}\rangle$ remains non-zero. We see that $\langle g_{\tilde{\q}=0,O}\rangle \approx U = 5$, which illustrates the strong inter-band scattering for mean-field electrons induced by $\tilde{\q}=0,\Q$ Goldstone modes. Fig. \ref{fig:gs} (c-d) shows the averaged scattering vertices in the rotating frame. From Fig. \ref{fig:gs} (d) we see that the inter-band scattering in the rotating frame indeed goes to zero as $\tilde{\q}\rightarrow 0,\Q$, and is now maximal at the magnetic Brillouin zone boundary. We also see that the intra-band scattering vertex $\langle g^R_{\tilde{\q}}\rangle$ is slightly enhanced in the rotating frame, and is almost identical (up to an overall scale factor) to the Goldstone dispersion $\omega_{\tilde{\q}}$ shown in Fig. \ref{fig:omega}.

We can also similarly define the averaged intra/inter-band scattering vertices $\langle f_{\tilde{\q},D/O}\rangle$ and $\langle f^R_{\tilde{\q},D/O}\rangle$ in the mean-field and rotating frame respectively. These quantities are shown in Fig. \ref{fig:fs}. Both the intra- and inter-band scattering vertices become smaller in the rotating frame, are maximal near the magnetic Brillouin zone boundary, but remain non-zero and order one as $\tilde{\q}\rightarrow 0,\Q$.

\subsection{Electron spectral function}\label{sec:spectfun}

In this last section, we go back to the original crystal momentum basis, and again assume that the AFM order is along the $x$-direction. The free fermion part of the effective action can then be written as

\begin{equation}
\begin{split}
    S_{\psi} = \int_0^\beta \mathrm{d}\tau & \sum_\k \sum_s \bar{\psi}_{\k,s}(\partial_\tau + \varepsilon_\k) \psi_{\k,s} \\
    &+ \Delta (\bar{\psi}_{\k,\uparrow}\psi_{\k+\Q,\downarrow} + \bar{\psi}_{\k+\Q,\downarrow}\psi_{\k,\uparrow})\,.
    \end{split}
\end{equation}
We will assume that these fermion fields are already in the rotated frame, such that their scattering with the Goldstone modes vanishes as $\q\rightarrow 0$. The components of the fermion Green's function in the rotated frame which are diagonal in the spin-$z$ indices are given by

\begin{equation}
\left[G^{R}_{\uparrow\uparrow}(i\omega,\k)\right]^{-1}  = \left[G^{R}_{\downarrow\downarrow}(i\omega,\k)\right]^{-1} = 
 i\omega - \varepsilon_\k - \frac{\Delta^2}{i\omega - \varepsilon_{\k+\Q}}\, .
\end{equation}
In experiments one obviously does not measure the electrons in the rotated frame. One measures the original physical fermions in the unrotated frame, which are given by $R^\dagger(\r)\bar{\psi}_{\r}$, with
\begin{equation}
    R(\r) = \exp\left(i[\varphi_y(\r)s^y + \varphi_z(\r)s^z]/2\right)\,,
\end{equation}
where $\varphi_y$ and $\varphi_z$ are the Goldstone boson fields. To encode the compactness of the Goldstone modes we need to write the quadratic Goldstone mode action in terms of a general SU$(2)$ matrix $R(\r)$ (and not in terms of $\varphi_y$ and $\varphi_z$). For this, it is easiest to start with the continuum version of the Goldstone mode action 
\begin{equation}
S_G =  \frac{1}{2}\int\mathrm{d}\tau\int\mathrm{d}^2\r\sum_{n=y,z} \chi(\partial_\tau\varphi_n)^2+ \rho (\nabla\varphi_n)^2 \,.\label{eq:SG}
\end{equation}
As a next step, we note that
\begin{equation}\label{Eq:88}
R^\dagger (-i\partial_\mu) R = \frac{1}{2}\sum_{n=y,z}\partial_\mu \varphi_n s^n + \mathcal{O}(\varphi^2)
\end{equation}
As a result, the following action agrees with Eq. \eqref{eq:SG} to lowest order in $\varphi$:
\begin{equation}\label{Eq:89}
S_G = \frac{1}{2}\int_{\tau,\r} \sum_{n=y,z} \chi\, \left[\text{tr}(R^\dagger i \partial_\tau R s^n)\right]^2 + \rho\, \left[\text{tr}(R^\dagger i \nabla R s^n)\right]^2
\end{equation}
This is the improved action which correctly incorporates the compactness of the order parameter manifold. Note that this action is invariant under
\begin{equation}\label{Eq:90}
R(\r) \rightarrow R(\r)e^{i\theta(\r)s^z}\,,
\end{equation}
with $\theta(\r)$ an arbitrary function. This is the U$(1)$ gauge invariance which is familiar from the CP$^1$ formulation of the AFM Goldstone action \cite{BookSachdev}. 

The Green's function of the physical fermions is given by
\begin{equation}
G_{s_1s'_1}(\tau,\r) = -\langle R^*_{s_2s_1}(\tau,\r)\psi_{s_2,\r}(\tau)R_{s'_2s'_1}(0,0)  \bar{\psi}_{s'_2}(0,0) \rangle_\beta \,,\nonumber
\end{equation}
with repeated indices summed over. As a first approximation, we ignore the interactions between the electrons and the Goldstone modes, in which case the Green's function factorizes:
\begin{equation}
G_{s_1s'_1}(\tau,\r) \approx -\langle R^*_{s_2s_1}(\tau,\r)R_{s'_2s'_1}(0,0)  \rangle\langle\psi_{s_2,\r}(\tau)\bar{\psi}_{s'_2}(0,0) \rangle_\beta \,.\nonumber
\end{equation}
This is expected to be a reasonable approximation as the fermions in the rotating frame decouple from the low-energy Goldstone modes. 

An approximate expression for the Goldstone propagator can be obtained from a large-N analysis of the CP$^1$ model \cite{PolyakovBook}, which leads to the following result \cite{Borejsza2004}:
\begin{equation}
\langle R^*_{s_2,s_1}(\tau,\r) R_{s'_2,s'_1}(0,0) 
 \rangle = - D(\tau,\r) \delta_{s_1s'_1}\delta_{s_2,s'_2}\,,
\end{equation}
where $D(\tau,\r)$ is the Fourier transform of
\begin{equation}\label{Eq:92}
D(i\nu,\q) = \frac{\chi^{-1}}{(i\nu)^2 -  c^2\q^2 - m^2}\, .
\end{equation}
Here $m^2$ is a small mass for the Goldstone modes generated by thermal fluctuations, which restore the symmetry in $2$ spatial dimensions due to the Hohenberg-Mermin-Wagner theorem.

Putting everything together, we find that the fermion Green's function is given by
\begin{equation}\label{Gphys}
\begin{split}
& G_{ss'}(i\omega,\k) = -\delta_{ss'}T\sum_{i\nu}\frac{1}{L_xL_y}\times \\
&\sum_\q\frac{\chi^{-1}}{(i\nu)^2 - \omega_\q^2 - m^2} 
2G^R_{\uparrow\uparrow}(i\omega-i\nu,\k-\q)
\end{split}
\end{equation}
This result has previously also been obtained by Borejsza and Dupuis \cite{Borejsza2004} starting from the rotating-frame mean-field approach motivated by Schultz \cite{Schulz1990}. To simplify Eq. \eqref{Gphys}, we first rewrite $G^R_{\uparrow\uparrow}$ as 
\begin{eqnarray}\label{eqn:BQP}
G^R_{\uparrow\uparrow}(i\omega,\k) & = & \frac{i\omega - \varepsilon_{\k+\Q}}{(i\omega - E_\k^+)(i\omega - E_\k^-)}  \\
& = & \frac{|u_{+}(\k)|^2}{i\omega -E_{\k}^+} + \frac{|u_{-}(\k)|^2}{i\omega -E_{\k}^-}\,,
\end{eqnarray}
where 
\begin{eqnarray}
E_\k^{\pm} & = & \frac{1}{2}\left(\varepsilon_\k + \varepsilon_{\k+\Q} \pm \sqrt{(\varepsilon_\k - \varepsilon_{\k+\Q})^2 + 4\Delta^2}\right)\,, \nonumber\\
|u_{\pm}(\k)|^2 & = &  \frac{1}{2}\left(1 \pm \frac{\varepsilon_{\k}-\varepsilon_{\k+\Q}}{\sqrt{(\varepsilon_\k - \varepsilon_{\k+\Q})^2 + 4 \Delta^2}} \right) \,.
\end{eqnarray}
It is now straightforward to perform the summation over $i\nu$ in Eq. \eqref{Gphys}, which gives
\begin{equation}
\begin{split}
G_{ss'}(i\omega,\k) = & \frac{\delta_{ss'}}{L_xL_y}\sum_\q \sum_{\sigma=\pm} \frac{|u_\sigma(\k-\q)|^2 }{\chi \tilde{\omega}_\q} \times \\
&\left(\frac{n(\tilde{\omega}_\q) + f(-E^\sigma_{\k-\q})}{i\omega - E^\sigma_{\k-\q}-\tilde{\omega}_\q} + \frac{n(\tilde{\omega}_\q) + f(E^\sigma_{\k-\q})}{i\omega - E^\sigma_{\k-\q}+\tilde{\omega}_\q}\right)\,,
\end{split}
\end{equation}
where $n(\omega)$ and $f(E)$ are respectively the Bose-Einstein and Fermi-Dirac distributions, and $\tilde{\omega}_\q = \sqrt{c^2\q^2 + m^2}$.

\begin{figure}
\includegraphics[scale=0.3]{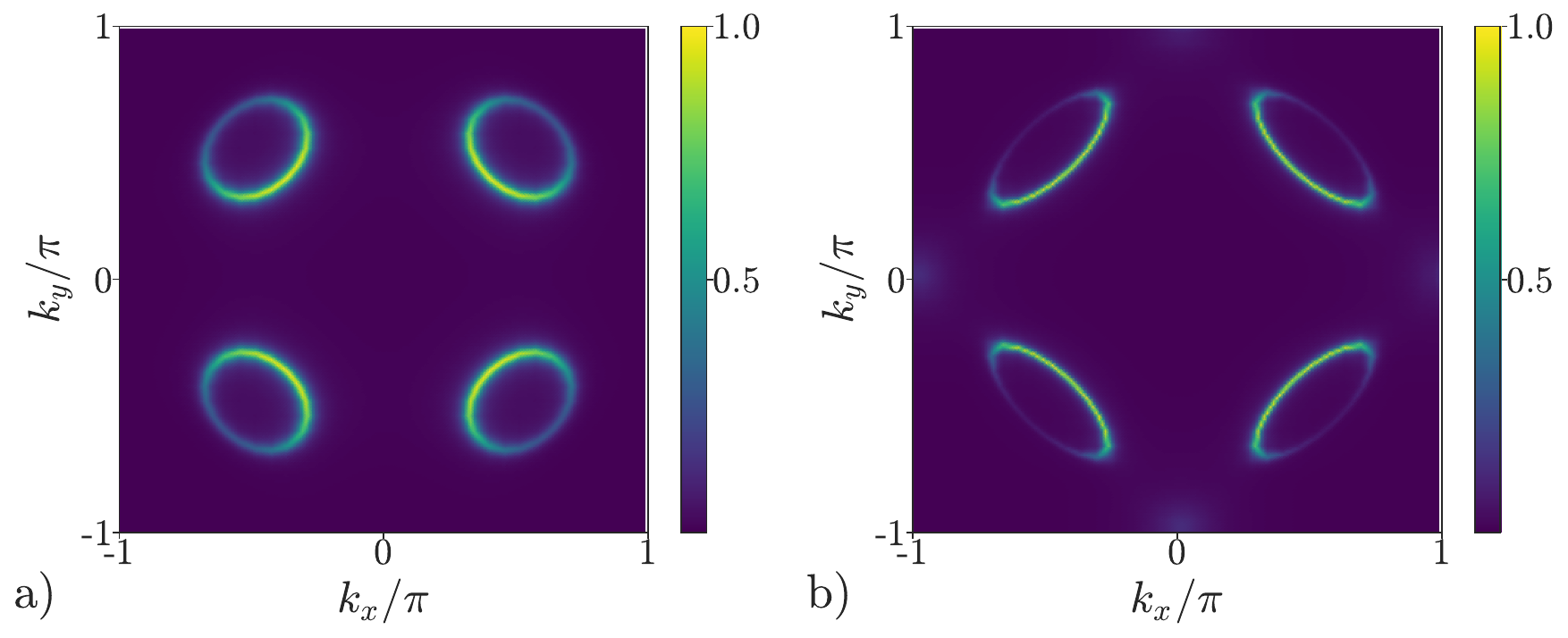}
\caption{Spectral weight $\mathcal{A}(\omega,\k)$ defined in Eq. \eqref{A} with $T = \epsilon=0.05$ and $m=0.002$, normalized such that the maximal value is one. (a) Spectral weight at frequency $\omega = -1.73$, using the mean-field value $\Delta \approx 1.93$. (b) Spectral weight at frequency $\omega = -0.53$, using a renormalized value $\Delta = 0.5$. In both (a) and (b), the total area contained in the pockets centered at $(\pm \pi/2,\pm \pi/2)$ is $\sim 10\%$ of the total Brillouin zone area. The results were obtained by numerically interpolating the mean-field results obtained on a $34\times 34$ momentum grid to a $170\times 170$ grid, using a bivariate spline interpolation.}\label{fig:A}
\end{figure}

The spectral function, defined for spin-rotation invariant systems as,
\begin{equation}
\mathcal{A}(\omega,\k) = \frac{2}{\pi} \text{Im}\,G_{\uparrow\uparrow}(\omega - i\epsilon,\k)
\end{equation}
is now easily obtained \cite{Borejsza2004}:
\begin{equation}
\begin{split}
\mathcal{A}&(\omega,\k) = \frac{2}{L_xL_y}\sum_\q \sum_{\sigma=\pm} \frac{|u_\sigma(\k-\q)|^2 }{\chi \tilde{\omega}_\q}\times\\
&\frac{\epsilon}{\pi}\left(\frac{n(\tilde{\omega}_\q) + f(-E^\sigma_{\k-\q})}{(\omega - E^\sigma_{\k-\q}-\tilde{\omega}_\q)^2 + \epsilon^2} + \frac{n(\tilde{\omega}_\q) + f(E^\sigma_{\k-\q})}{(\omega - E^\sigma_{\k-\q}+\tilde{\omega}_\q)^2 + \epsilon^2}\right)\,.\label{A}
\end{split}
\end{equation}
In Fig. \ref{fig:A} (a) we show the numerically obtained spectral weight $\mathcal{A}(\omega,\k)$ at frequency $\omega = -1.73$. To obtain this result, we have kept a small non-zero $\epsilon = T = 0.05$ to smear out the numerical results obtained on a finite-size discrete momentum grid. We see that at this negative frequency, the spectral weight is located along contours demarcating the boundaries of `hole pockets' centered at $(\pm \pi/2,\pm \pi/2)$. The part of the contour oriented towards the center of the Brillouin zone is brighter than the backside facing $(\pi,\pi)$ as a result of the coherence factors $|u_\sigma(\k)|^2$ in Eq. \eqref{A}. 

Fluctuations will cause the AFM order to decrease. To illustrate the effect of a reduced order in the rotating frame on the spectral function, we show the spectral weight at $\omega = -0.53$ obtained using a smaller $\Delta = 0.5$ in Fig. \ref{fig:A} (b). Note that both in Fig. \ref{fig:A} (a) and (b), the area inside the `hole pockets' is $\sim 10 \%$ of the total Brillouin zone area. From Fig. \ref{fig:A} (b) we see that reducing $\Delta$ elongates the contours of high spectral weight in the directions toward $(0,\pm \pi)$ and $(\pm \pi,0)$. Also the spectral weight on the backside of the contours is further reduced.

Before concluding we want to reiterate that the spectral weights shown in Figs. \ref{fig:A} (a-b) are those of a thermally disordered state, with no long-range AFM order. Nevertheless, the spectral weight retains important features of the $T=0$ AFM state, and is strikingly different from the spectral weight of the conventional Fermi liquid at small $U$.

\section{Example II: spin-spiral order in the three-band model}\label{sec:ex2}

Having discussed the anti-ferromagnetic Mott insulator in the Hubbard model in the previous section, we now turn to the slightly more involved example of spin spiral order in the hole-doped three-band Hubbard model \cite{Emery87}.

\subsection{Mean-field state and Goldstone mode energies}

\begin{figure}
    \centering
   \includegraphics[scale=0.9,trim = 7.7cm 11.2cm 8.2cm 11.4cm, clip]{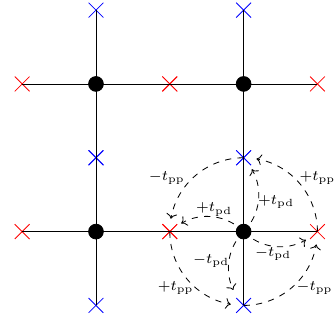}
    \caption{Hopping processes in the three-band model. The copper $d_{x^{2}-y^{2}}$ orbitals are denoted by the black dots while the Oxygen $2p_{x}$ and $2p_{y}$ orbitals are denoted by the red and blue crosses respectively. The notation for the hopping amplitudes is the same as in Eq. \eqref{eq:SpiralHamiltonian}.}
    \label{fig:3band}
\end{figure}

The three-band model is described by the following Hamiltonian:
\begin{equation}\label{eq:SpiralHamiltonian}
    \begin{split}
        & H = \sum_{i} \epsilon_{d} c^{\dagger}_{1i}c_{1i} + \epsilon_{p} \left( c^{\dagger}_{2i}c_{2i} + c^{\dagger}_{3i}c_{3i}  \right) \\
        & + \sum_{\langle ij \rangle} t^{ij}_{pd} \left( c^{\dagger}_{1i}c_{2j} + c^{\dagger}_{1i}c_{3j} + h.c.\right) 
         + \sum_{\langle ij \rangle} t^{ij}_{pp} \left( c^{\dagger}_{2i}c_{3j} +  h.c.\right)     \\
        & + \sum_{i} U_{d} n_{1i\uparrow}n_{1i\downarrow} + U_{p} \left( n_{2i\uparrow}n_{2i\downarrow} + n_{3i\uparrow}n_{3i\downarrow} \right)  \,,
    \end{split}
\end{equation}
where $a=1,2,3$ respectively refers to the Copper $d_{x^2-y^2}$, Oxygen $2p_x$ and $2p_y$ orbitals, $s$ denotes the spin, and $i,j$ label the unit cells.
Note that following most of the literature on this model, we have formulated the Hamiltonian in terms of the hole degrees of freedom. The first two lines in Eq. \eqref{eq:SpiralHamiltonian} represent a potential energy difference for the Copper and Oxygen orbitals, and nearest and next-nearest neighbour hopping (in these terms, the spin summation is implicit). Note that the signs of the hopping parameters $t^{ij}_{pd}$ and $t^{ij}_{pp}$ depend on the relative orientation of the sites $i$ and $j$. See Fig. \ref{fig:3band} for a graphical representation of the different hopping processes in the three-band model, and how the corresponding signs depend on the orientation. The last line in Eq. \eqref{eq:SpiralHamiltonian} contains the on-site Hubbard interactions for electrons in the Copper and Oxygen orbitals.

An important parameter in the three-band model is the charge-transfer parameter $\delta = \epsilon_{p}-\epsilon_{d} > 0$. Previous works \cite{Chiciak2018,Chiciak2020} have investigated the groundstate properties of the three-band model for a range of values of $\delta$ using different numerical methods, including Hartree Fock, and identified a parameter regime with spin-spiral order. To make contact with the results of \cite{Chiciak2018}, we choose the following values for the parameters of the three-band model: $U_{d}=8.4,U_{p}=2.0,\epsilon_{d} = -7.6, \epsilon_{p}=-6.1, |t^{ij}_{pd}|=1.2, |t^{ij}_{pp}|=0.7$. Note that for these parameter values the charge-transfer parameter is $\delta = 1.5$. At half filling, these parameters lead to an insulating AFM ground state, with orbital hole densities $n_1 \approx 0.46$,$n_2 = n_3 \approx0.27$. Upon hole doping, the anti-ferromagnet becomes a circular spin spiral with wave vector $\Q = (\pi,\pi-2\pi\eta)$ or $\Q = (\pi-2\pi\eta,\pi)$, where $\eta$ is the incommensurability of the spiral. In this work we investigate the ground state at a hole doping of $1/8$. 

Let us assume without loss of generality that the spiral order is in the XY plane. In this case, the spin expectation value is given by
\begin{equation}
    \langle \mathbf{S}_{i} \rangle = \Delta \left( \cos(\Q \cdot \mathbf{r}_{i})\mathbf{\hat{x}}+\sin(\Q \cdot \mathbf{r}_{i})\mathbf{\hat{y}} \right)\,,
\end{equation}
where $\mathbf{S}_{i} = \frac{1}{2}\sum_{a,s,s'} c^{\dagger}_{j,s,a} \mathbf{\sigma}_{ss'} c_{j,s',a}$. Similar to the AFM, the circular spiral order breaks both SU$(2)$ spin symmetry and translation symmetry, but is invariant under the modified translation symmetry
\begin{equation}
    T'_{\r} = e^{i\r\cdot \Q s^z/2} T_{\r}\,,
\end{equation}
which allows us to define a conserved pseudo-momentum $\tilde{\mathbf{k}}$ as in Eq.\eqref{75}.

The self-consistent Hartree-Fock mean-field Hamiltonian can then be written as
\begin{equation}\label{HHF_Emery}
H_{HF} = \sum_{\tilde{\k}} c^\dagger_{\tilde{\k}} \left(\begin{matrix} \varepsilon_{\tilde{\k},\uparrow} & \Delta_{\tilde{\k}} \\ \Delta_{\tilde{\k}} & \varepsilon_{\tilde{\k},\downarrow}\end{matrix}\right) c_{\tilde{\k}}\, ,
\end{equation}
where $c^\dagger_{\tilde{\k}} = (c^\dagger_{1,\tilde{\k},\uparrow},c^\dagger_{2,\tilde{\k},\uparrow},c^\dagger_{3,\tilde{\k},\uparrow},c^\dagger_{1\tilde{\k},\downarrow},c^\dagger_{2\tilde{\k},\downarrow},c^\dagger_{3\tilde{\k},\downarrow})$ and 
 $\Delta_{\tilde{\k}} = \text{diag(}\Delta_{1,\tk},\Delta_{2,\tk},\Delta_{3,\tk}\text{)}$. Here, $\varepsilon_{\tilde{\k}}$ are not the bare band energies since the contribution from the Hartree term is not just a simple overall energy shift, but rather, each orbital gets shifted by a different amount, corresponding to the orbital's hole density. 

 To solve the HF self-consistency equations, we have applied the following procedure. First, we used restricted HF with conserved pseudo-momentum, and minimized the energy for different values of $\Q$. Even though the optimal $\Q$ varies slightly with system size, it was consistently found to be $\Q\approx(\pi,0.80\pi)$ (assuming spiral order along the $y$-direction). As a next step we used the optimal solution of the restricted HF as an initial seed for completely unrestricted HF, which did not assume any translational symmetry. The converged unrestricted HF yielded ground states with the same spiral order as the restricted HF, but with a small $\delta n /n \sim 10^{-5}-10^{-3} $ charge density modulation, which may be a result of the incommensurability of the exact optimal $\Q$ with the finite system size. The difference in energy between the spiral states with and without charge density modulation was found to decrease with system size. Our results are thus consistent with a circular spin spiral state with $\Q = (\pi,\pi - 2\pi \eta)$ and $\eta \approx 0.10$, in agreement with Refs. \cite{Chiciak2018,Cui2020}. The spin hybridization parameters in the mean-field Hamiltonian were found to be $\left(\Delta_{1,\tk},\Delta_{2,\tk},\Delta_{3,\tk} \right) = \left(\Delta_1,\Delta_2,\Delta_3 \right)\approx \left(1.31,0.00,0.013 \right)$.

\begin{figure*}
    \hspace{-0.5cm}\includegraphics[scale=0.6]{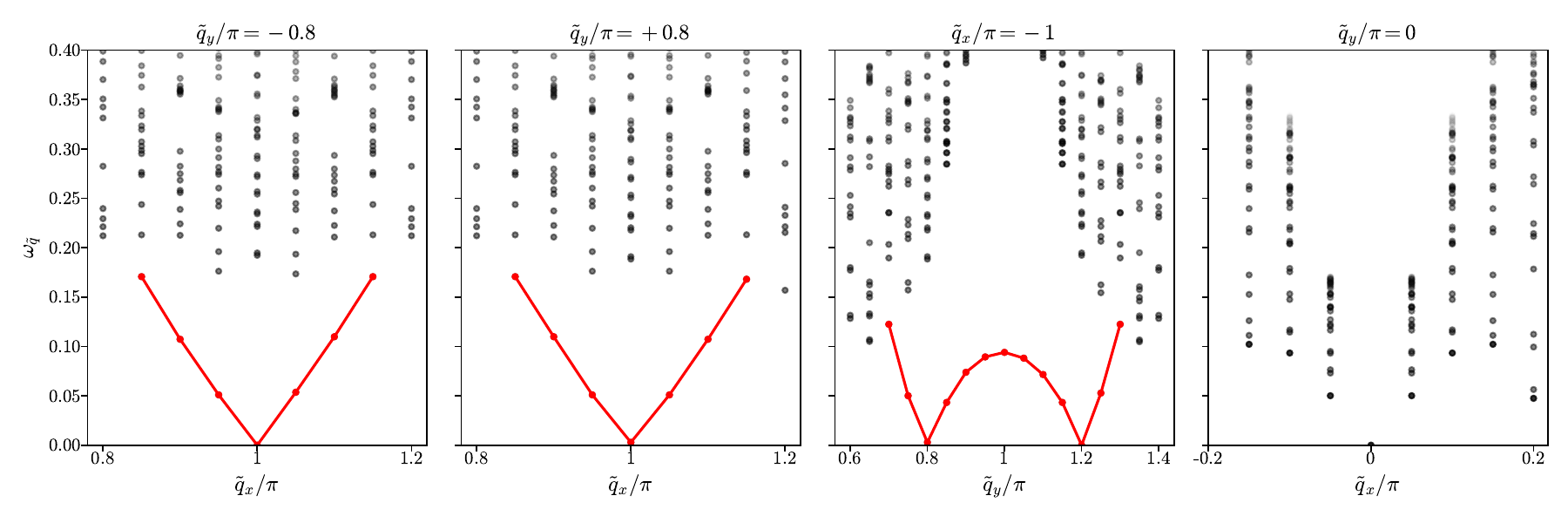}
    \caption{Plots of the collective modes of the spin spiral state on a $40\times 40$ grid, across different cuts of the Brillouin Zone. The red points show the Goldstone modes where they can be distinguished from the particle-hole continuum . The linear low-energy dispersion of the Goldstone modes near $\tilde{\q}=\pm \Q = (\pi,\pm 0.8\pi)$ is clear in the first three figures. In the fourth and rightmost figure, the Goldstone mode at $\tilde{\q}=(0,0)$ is dissolved in the particle-hole continuum.}
    \label{fig:CollectiveModeSpiral}
\end{figure*}

The spiral order has three broken symmetry generators, and three Goldstone modes \cite{Kampf1996,Zhou1995}. Two of the modes are associated with rotating the plane of the spiral order (out-of-plane modes), and one mode corresponds to in-plane rotations. For a $XY$ spiral, the out-of-plane modes are related to the broken $s^x,s^y$ generators while the in-plane mode is related to the broken $s^z$ generator. We can understand the location of the Goldstone modes in the pseudo-momentum frame by relating them to the original frame, where they are all at $\q=(0,0)$. The in-plane mode remains at zero, $\Tilde{\q}=(0,0)$, while the two out-of-plane modes go to $\Tilde{\q}= \pm \Q$. We obtain the collective mode spectrum by numerically solving Eq.\eqref{collmode} in the pseudo-momentum basis. This was first done over the entire pseudo-momentum Brillouin zone on a $24\times24$ grid, and subsequently on a larger $40\times40$ grid near the locations of the Goldstone modes. The energies for the low-energy collective modes are shown in Fig.\ref{fig:CollectiveModeSpiral}, near $\Tilde{\q}=(0,0)$ and $\Tilde{\q}=\pm \Q$. We see that the Goldstone modes near $\pm \Q$, which correspond to the out-of-plane fluctuations, lie outside the particle-hole continuum and hence will not be damped. The Goldstone mode near $\tilde{\q}$, on the other hand, does lie in the particle-hole continuum, and hence will be Landau damped. These observations agree with Refs. \cite{Kampf1996,Zhou1995,Bonetti2022_2}. The velocities of the out-of-plane Goldstone modes are anisotropic, and from a linear fit we find $c_{x} \approx 0.32$ and $c_{y} \approx 0.27$.

\subsection{Electron spectral function}
Working in the original crystal momentum basis, and taking the spiral order to be in the XY plane, the fermion action is given by form
\begin{equation}
\begin{split}
    & S_{\psi} =  \int_0^\beta \mathrm{d}\tau  \sum_\k \sum_{s,a,b} \bar{\psi}_{\k,s,a}(\delta_{ab}\partial_\tau + \varepsilon_{\k,ab}) \psi_{\k,s,b} \\
    &+ \sum_{a}\Delta_{a} (\bar{\psi}_{\k,\uparrow,a}\psi_{\k+\Q,\downarrow,b} + \bar{\psi}_{\k+\Q,\downarrow,b}\psi_{\k,\uparrow,a})\,.
    \end{split}
\end{equation}
As before, we assume that these fermion fields are already in the rotated frame. In a basis $(\psi_{\k,\uparrow},\psi_{\k+\Q,\downarrow})$, suppressing the orbital indices, the rotated-frame Green's function is given by
\begin{equation}
    [G^{R}(i\omega,\k)]^{-1} = \begin{pmatrix}
        i\omega \mathbb{1} - \varepsilon_{\k} & -\Delta\\ -\Delta & i\omega \mathbb{1} - \varepsilon_{\k+\Q}\\
    \end{pmatrix} \, ,
\end{equation}
where $\Delta,\varepsilon_{\k}$ are $3\times3$ matrices with orbital indices, with values as described in the previous section. 

The spin-diagonal part of the Green's function can be written as
\begin{equation}
    G^{R\hspace{0.05cm}ab}_{\hspace{0.25cm}ss}(i\omega,\k) = \sum_{\alpha} \frac{u^{as}_{\alpha}(\k)u^{bs \hspace{0.01cm}*}_{\alpha}(\k)}{i\omega-E_{\alpha}(\k)} \,,
\end{equation}
which is the equivalent of Eq. \eqref{eqn:BQP} in the one-band Hubbard model case. Note that for the spiral order, it no longer holds that $G^{R}_{\uparrow\uparrow}(i\omega,\k) = G^{R}_{\downarrow\downarrow}(i\omega,\k)$. 

The microscopic fermions $R^{\dagger}(\r)\bar{\psi}_{\r}$ are obtained by acting with
\begin{equation}
    \begin{split}
    R(\r) &= \exp\left(i[\varphi_y(\r)\mathbb{1}\otimes s^y + \varphi_z(\r)\mathbb{1} \otimes s^z]/2\right) \\ 
    &= \exp\left(i[\varphi_x(\r)\bar{s}^x+\varphi_y(\r)\bar{s}^y + \varphi_z(\r)\bar{s}^z]/2\right) \,,
\end{split}
\end{equation}
on the fermions in the rotating frame.
Here we have defined $\bar{s}^{i} = \mathbb{1}\otimes s^{i}$ are the correct generators which act trivially on the orbital indices. The quadratic Goldstone action is then given by
\begin{equation}\label{eq:SGspiral}
\begin{split}
S_G =  \frac{1}{2}\int\mathrm{d}\tau & \int\mathrm{d}^2\r \sum_{n=x,y } \left( \chi^{\perp}(\partial_\tau\varphi_n)^2+ \sum_{i=x,y}\rho^{\perp}_i (\partial_i \varphi_n)^2 \right) \\
& + \chi^{\Box}(\partial_\tau\varphi_z)^2+  \sum_{i=x,y}\rho^{\Box}_i (\partial_i\varphi_z)^2 \,,
\end{split}
\end{equation}
where the out-of-plane and in-plane parameters are denoted by $\perp$ and $\Box$. Note that terms involving $\partial_{i}\varphi \partial_{j}\varphi$ with $i\neq j$ are forbidden by the reflection symmetry of the spiral state. 

By again using Eq. \eqref{Eq:88}, we obtain the following non-linear sigma action for the Goldstone modes:
\begin{equation}
\begin{split}
 S_G = & \frac{1}{2}\int_{\tau,\r} \sum_{n=x,y} \chi^\perp \left[ \text{tr}(  R^\dagger i\partial_\tau R s^n)\right]^2 + \rho_j^\perp \left[ \text{tr}(  R^\dagger i\partial_jR s^n)\right]^2 \\
& + \chi^\Box \left[ \text{tr}(  R^\dagger i\partial_\tau R s^z)\right]^2 + \rho_j^\Box \left[ \text{tr}(  R^\dagger i\partial_jR s^z)\right]^2\,,
\end{split}
\end{equation}
where the sum over $j=x,y$ is implicit. A large-N analysis of the non-linear sigma model \cite{Bonetti2022}, leads to the following result for the Goldstone propagator:
\begin{equation}\label{eq:goldstonepropagator}
\langle R^{c,a\hspace{0.1cm} *}_{s_2,s_1}(\tau,\r) R^{d,b}_{s'_2,s'_1}(0,0)  \rangle = - D(\tau,\r) \delta_{s_1s'_1}\delta_{s_2s'_2}\delta_{ad}\delta_{cb}\,, 
\end{equation}
where $D(\tau,\r)$ is the Fourier transform of
\begin{equation}\label{eq:goldstonepropagator2}
    D(i\nu,\q)  = \frac{\hspace{0.3cm}(\chi^{\perp})^{-1}}{(i\nu)^2 - \tilde{\omega}_{\q}^2 }\, .
\end{equation}
Here, $\tilde{\omega}_{\q}= \sqrt{\rho^{\perp}_{\alpha}q_{\alpha}^2/\chi^{\perp} + m^{2}}$ is the Goldstone mode dispersion, including the exponentially small (in $T$) spin-gap mass generated by the thermal fluctuations at non-zero temperature, as determined via the saddle point equations. The velocities $c_{\alpha} = \sqrt{\rho^{\perp}_{\alpha}/\chi^{\perp}}$ are obtained from the collective mode spectrum.

Finally, we express the fermion Green's function as 
\begin{equation*}
    \begin{split}
        G^{ab}_{s_1s'_1}(\tau,\r) &= -\langle R^{ca \hspace{0.1cm}*}_{s_2s_1}(\tau,\r) \psi^{c}_{s_2}(\tau,\r)  \bar{\psi}^{d}_{s_2'}(0,0) R^{db}_{s'_2s'_1}(0,0) \rangle_\beta  \\
        & \approx -\langle R^{ca \hspace{0.1cm}*}_{s_2s_1}(\tau,\r) R^{db}_{s'_2s'_1}(0,0)  \rangle\,   G^{\text{R}\hspace{0.05cm} cd}_{ \hspace{0.23cm} s_2 s_2'} (\tau,\r) 
    \end{split}
\end{equation*}
where we have again ignored the interaction between the Goldstone modes and the electrons in the rotating frame, so that the Green's function factorizes. Going to momentum space, and using Eqs. \eqref{eq:goldstonepropagator} and \eqref{eq:goldstonepropagator2}, we find that the fermion Green's function is given by
\begin{equation}
\label{SpiralGreensfunction}
    \begin{split}
        G^{ab}_{ss'}(i\omega,\k) & = -\delta_{ss'} T \sum_{i\nu} \frac{1}{N_x N_y} \times \\
         & \sum_{\q} \frac{\hspace{0.3cm}(\chi^{\perp})^{-1}}{(i\nu)^2 - \tilde{\omega}_{\q}^2 } \sum_{\sigma,\alpha}  \frac{u^{a,\sigma }_\alpha(\k-\q)u^{b,\sigma\hspace{0.05cm} *}_\alpha(\k-\q)}{i(\omega-\nu)-E_{\k-\q,\alpha}}\, ,
    \end{split}
\end{equation}
from which we obtain the spectral weight.

\begin{figure}
\includegraphics[scale=0.565]{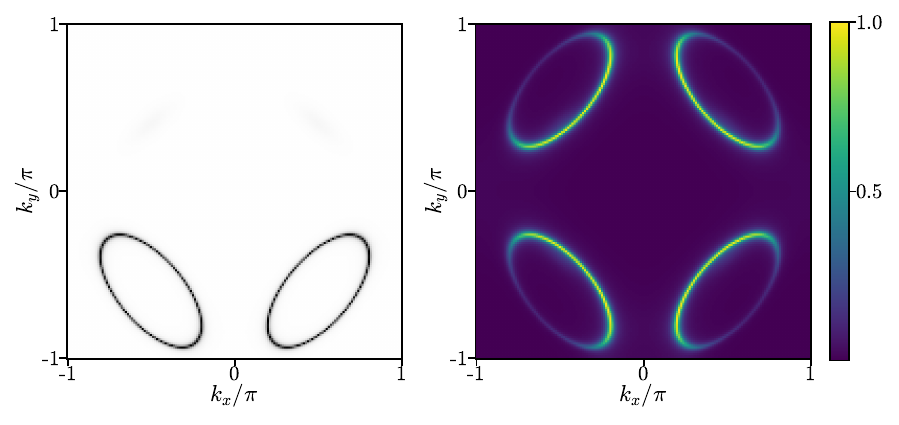}
\caption{Plot of the mean-field Fermi surfaces (left) and normalized spectral weight $\mathcal{A}(\omega = 0 ,\k)$ at the Fermi energy of the fermions $f^\dagger_\r(\theta^* = 0.27\pi)$ as defined in \eqref{eq:PhysicalCreationOperator} (right), using $T=\epsilon=0.05$. The results were obtained by interpolating the mean-field results on a $40 \times 40$ momentum grid to a $160 \times 160$ grid.}
\label{fig:Spiral_A_3band}
\end{figure}

In Fig \ref{fig:Spiral_A_3band}(a) we show the Fermi surfaces of the spiral state mean-field band structure in the pseudo-momentum Brillouin zone. The band spectrum clearly breaks $C_{4}$ symmetry, in contrast to the AFM case, but retains a reflection symmetry about $x$. The spectral weight of the physical electrons involves a spin sum, as seen in Eq. \eqref{SpiralGreensfunction}, which adds two copies of the Fermi levels of Fig \ref{fig:Spiral_A_3band}(a), with a relative shift of $\Q$ and the associated coherence factors. This restores the reflection symmetry about $y$, but not the $C_{4}$ symmetry. To illustrate this, we have calculated the spectral weight of the following fermions:
\begin{equation}
\label{eq:PhysicalCreationOperator}
        f^{\dagger}_{\r}(\theta) = \cos \theta \,c^{\dagger}_{1,\r} +
        \frac{\sin \theta}{2}\left( c^{\dagger}_{2,\r} - c^{\dagger}_{3,\r} - c^{\dagger}_{2,\r-x} + c^{\dagger}_{3,\r-y}  \right) \, ,
\end{equation}
where the combination of orbitals in this operator is chosen to obey all the point group symmetries of the three-band model. We have optimized the spectral weight at the Fermi energy as a function of $\theta$, and found that the maximal spectral weight occurs at $\theta = \theta^* \approx 0.27\pi$. In Fig. \ref{fig:Spiral_A_3band}(b) we plot the spectral weight of $f^\dagger_\r(\theta^*)$, which is indeed reflection symmetric, but breaks $C_4$. Similar to the AFM case discussed above, the highest spectral weight is along four contours with a suppressed backside due to the coherence factors $|u^{a}_{\alpha}(\k)|^{2}$. However, compared to the AFM, the centers of the high spectral-weight contours are shifted along $k_{y}$. The fraction of the Brillouin zone area contained in the Fermi surface is equal to the number of doped holes per unit cell relative to half filling. However, since the incommensurate spiral order has only two hole pockets compared to the AF order which has four, the area enclosed in each pocket is twice as large. The experimental implications of this were previously discussed in Ref. \cite{Eberlein2016}.

\section{Conclusions}

We have shown that a naive application of mean-field theory + RPA produces scattering vertices between electrons and Goldstone modes which do not vanish in the long-wavelength limit. However, fermions which live in a frame that locally follows the order parameter fluctuations do decouple from $\q=0$ Goldstone modes. This has important consequences for electron correlation functions, which by making use of the rotating frame naturally reflect the properties of thermally disordered systems which nevertheless retain features associated with a non-zero magnitude for the order parameter. 

Although we have illustrated our formalism only for square-lattice Hubbard and three-band models, its applicability is far more general. In particular, it can also be used to study the broken-symmetry states in moir\'e materials, such as e.g. the Incommensurate Kekul\'e Spiral (IKS) state \cite{IKS,GlobalPD,IKSDMRG} which is observed experimentally in both magic-angle twisted bilayer \cite{Nuckolls2023} and trilayer \cite{Kim2023} graphene. The IKS order is in fact very similar to the circular spin-spiral order that we have studied in the three-band model in Sec. \ref{sec:ex2}. We leave these applications for near-future work.

\emph{Acknowledgements --} N.B. would like to thank Patrick Ledwidth for helpful discussions about the manuscript, and Steve Kivelson for bringing Ref. \cite{Ramakrishnan1989} to our attention. This work was supported by a Leverhulme Trust International Professorship grant [number LIP-202-014]. For the purpose of Open Access, the authors have applied a CC BY public copyright licence to any Author Accepted Manuscript version arising from this submission. The research was also in part supported by the National Science Foundation under Grant No. NSF PHY-1748958. N.B. was supported by a University Research Fellowship of the Royal Society, and thanks \'Ecole Normale Sup\'erieure Paris, where part of this work was completed, for hospitality. Y.H. acknowledges support from the European Research Council (ERC) under the European Union Horizon 2020 Research and Innovation Programme (Grant Agreement Nos. 804213-TMCS)

\begin{appendix}

 \section{Solution of the Bethe-Salpeter equation and its properties}\label{app:BS}

 In this appendix we elaborate on the Bethe-Salpeter equation, its solutions, and their properties. We will closely follow the seminal works \cite{Thouless1960,Thouless1961}. To ease the presentation, we will consider a general Hartree-Fock action of the form

 \begin{equation}
     \int_0^{1/T}\mathrm{d}\tau\, \sum_\alpha \bar{\psi}(\partial_\tau + E_\alpha)\psi_\alpha + \frac{1}{2}\sum_{\alpha\beta\lambda\sigma}V_{\alpha\lambda,\beta\sigma} \bar{\psi}_\alpha\bar{\psi}_{\lambda}\psi_\sigma \psi_\beta\,,
 \end{equation}
 where the Greek indices run over the Hartree-Fock single-particle states. 
 
 As explained in the main text, the RPA collective mode propagator is obtained by summing the infinite set of diagrams in Eq.~\eqref{G2}. Summing this infinite series is equivalent to solving the Bethe-Salpeter equation, whose graphical representation we repeat here:

\begin{equation}
    \includegraphics[scale=0.7]{G2_BS}
\end{equation}
Written out explicitly, the Bethe-Salpeter (BS) equation is
\begin{widetext}
\begin{eqnarray}
G_2(\alpha,\beta;\lambda,\sigma;i\nu) & = & T \sum_{\omega_n} \left(-\delta_{\beta\sigma}\delta_{\lambda\alpha}\frac{1}{i(\omega_n-\nu) - E_\beta}\frac{1}{i\omega_n - E_\alpha} \right) \\
 & & + \sum_{\mu\nu}V_{\alpha\mu,\beta\nu} G_2(\nu,\mu;\lambda,\sigma;i\nu) T \sum_{\omega_n}\frac{1}{i(\omega_n-\nu) - E_\beta}\frac{1}{i\omega_n - E_\alpha}  \nonumber \\
 & & - \sum_{\mu\nu}V_{\alpha\mu,\nu\beta} G_2(\nu,\mu;\lambda,\sigma;i\nu) T \sum_{\omega_n}\frac{1}{i(\omega_n-\nu) - E_\beta}\frac{1}{i\omega_n - E_\alpha}  \nonumber \\ 
  & = & -\frac{f(E_\beta)-f(E_\alpha)}{i\nu - E_\alpha + E_\beta}\left(\delta_{\alpha\lambda}\delta_{\beta\sigma} - \sum_{\mu\nu}\left(V_{\alpha\mu,\beta\nu} - V_{\alpha\mu,\nu\beta} \right) G_2(\nu,\mu;\lambda,\sigma;i\nu)\, \right),\nonumber
\end{eqnarray}
where $f(E)$ is the Fermi-Dirac distribution. By going to zero temperature we find that $G_2$ satisfies the following equation:

\begin{eqnarray}
\sum_{\mu,\nu}\left( (n_\beta-n_\alpha)(E_\alpha - E_\beta - i\nu)\delta_{\alpha\nu}\delta_{\beta\mu} + V_{\alpha\mu,\beta\nu} - V_{\alpha\mu,\nu\beta}\right) G_2(\nu,\mu;\lambda,\sigma;i\nu) =  \delta_{\alpha\lambda}\delta_{\beta\sigma}\, ,
\end{eqnarray}
\end{widetext}
where $n_\alpha \in \{0,1\}$ are the fermion occupation numbers. Again adopting the notation where $i,j,k,l$ represent occupied states, and $m,n,o,p$ unoccupied states, we can rewrite the BS equation as

\begin{equation}\label{matrixSD}
\left(\begin{matrix} A - i\nu \mathds{1} & B \\ B^* & A^* + i\nu\mathds{1}\end{matrix}\right)\left(\begin{matrix}  G^{(1)}(i\nu) & G^{(2)}(i\nu) \\ G^{(3)}(i\nu) & G^{(4)}(i\nu)\end{matrix}\right) = \left(\begin{matrix}\mathds{1} & 0 \\ 0 & \mathds{1} \end{matrix}\right)\, ,
\end{equation}
where we have defined

\begin{eqnarray}
G^{(1)}_{mi,nj}(i\nu) & = & G_2(m,i;n,j;i\nu) \\
G^{(2)}_{mi,nj}(i\nu) & = & G_2(m,i;j,n;i\nu) \\
G^{(3)}_{mi,nj}(i\nu) & = & G_2(i,m;n,j;i\nu) \\
G^{(4)}_{mi,nj}(i\nu) & = & G_2(i,m;j,n;i\nu)\, ,
\end{eqnarray}
and the matrices $A$ and $B$ are given by

\begin{eqnarray}
A_{mi,nj} & = & (E_m - E_i) \delta_{ij}\delta_{mn} + V_{mj,in} - V_{mj,ni} \label{eqA}\\
B_{mi,nj} & = & V_{mn,ij} - V_{mn,ji} \label{eqB}
\end{eqnarray}
The complex conjugate in the lower rows in Eq. \eqref{matrixSD} comes from the following property of the interaction potential:

\begin{equation}
V^*_{\alpha\lambda,\beta\sigma} = V_{\beta\sigma,\alpha\lambda}\, ,
\end{equation}
which is a simple consequence of hermiticity of the Hamiltonian. The anti-symmetry of the fermion creation and annihilation operators implies that the interaction potential also satisfies

\begin{equation}
V_{\alpha\lambda,\beta\sigma} = V_{\lambda\alpha,\sigma\beta}\, ,
\end{equation}
from which it follows that $A$ is hermitian, and $B$ is symmetric. As a result, the following matrix

\begin{equation}
    H_B = \left(\begin{matrix}A & B \\ B^* & A^* \end{matrix}\right)
\end{equation}
is hermitian. It also has a particle-hole symmetry:

\begin{equation}\label{PHapp}
    H_B^* = XH_BX\,,\text{ with } X = \left(\begin{matrix}0 & \mathds{1} \\ \mathds{1} & 0 \end{matrix}\right)\,.
\end{equation}
Note that this is a true particle-hole transformation for the excitations on top of the HF state, as $X$ maps from $(i,m)$ space to $(m,i)$ space, so it maps electron excitations to hole excitations and vice versa.

The solution to Eq. \eqref{matrixSD} is given by

\begin{equation}
    G_2(i\nu) = \left(ZH_B - i\nu \mathds{1} \right)^{-1}Z\,,
\end{equation}
where

\begin{equation}
    Z = \left(\begin{matrix}\mathds{1} & 0 \\ 0 & -\mathds{1} \end{matrix}\right)\, .
\end{equation}
The matrix $ZH_B$ satisfies $X(ZH_B)^*X = -ZH_B$, so all its non-zero eigenvalues come in pairs $\pm \omega$. Here we used that $\omega$ is real whenever $H_B$ is a positive matrix, which is the condition for the Hartree-Fock state to be a variational energy minimum \cite{Thouless1960,Thouless1961}). Let us now assume that $ZH_B$ has no zero modes, which will generically be true if we add for example a very small symmetry breaking field to the Hamiltonian which gaps the Goldstone modes. In the absence of zero modes, the eigenvectors of $ZH_B$ span a complete basis \cite{Thouless1961}, which means that the matrix $S$ whose columns consist of these eigenvectors is invertible. We can therefore write $ZH_B$ as

\begin{equation}\label{geneig}
    ZH_B = S(Z_\eta\Omega)S^{-1}\, ,
\end{equation}
where $\Omega$ is a positive diagonal matrix, and $Z_\eta\Omega$ is a diagonal matrix containing the eigenvalue pairs $\pm \omega$ ($Z_\eta$ is thus a diagonal matrix with $\pm 1$ on the diagonal). The particle-hole symmetry in Eq. \eqref{PHapp} implies that if $S_{(\alpha\beta),s}$ is an eigenvector with eigenvalue $\omega_s$, then $[XS^*]_{(\alpha\beta),s}$ is an eigenvector with eigenvalue 
$-\omega_s$. 

Using the hermiticity of $H_B$ and Eq. \eqref{geneig}, we can write

\begin{eqnarray}
    [S^\dagger H_B S]_{ss'} & = & [S^\dagger Z S]_{ss'} [Z_\eta\Omega]_{s's'} \label{SHS1}\\
    & = & [Z_\eta\Omega]_{ss} [S^\dagger Z S]_{ss'} \label{SHS2}
\end{eqnarray}
Comparing Eqs. \eqref{SHS1} and \eqref{SHS2}, we conclude that if $[Z_\eta\Omega]_{s's'} \neq [Z_\eta\Omega]_{ss}$, then $[S^\dagger Z S]_{ss'} = 0$. Moreover, as $H_B$ is positive if the Hartree-Fock state is a variational energy minimum, which we assume to be the case here, it follows that 

\begin{equation}\label{SZSsgn}
\text{sgn}([S^\dagger Z S]_{ss}) = [Z_\eta]_{ss} \text{ if } \Omega_{ss}=\omega_s \neq 0. 
\end{equation}
As we assume the Goldstone modes to be gapped, and hence that there are no zero modes, we can normalize the columns of $S$ (i.e. the eigenvectors of $ZH_B$) such that the following equation holds:

\begin{equation}\label{SZS}
    S^\dagger Z S = Z_\eta\, ,
\end{equation}
Equation \eqref{SZS} is Eq. \eqref{symplnorm} in the main text. To see this, note that if we restore the momentum labels, we have that $[S(\q)]_{(\k\alpha\beta),s} = \varphi^s_{\q,\alpha\beta}(\k)$, i.e. the columns of $S$ (which is block-diagonal in momentum $\q$) are labeled by $s$ and correspond to the collective mode wavefunctions. The diagonal elements of $Z_\eta$ are the signs $\eta_{\q,s}$ defined in the main text. 

Using the eigenbasis of $ZH_B$, normalized such that Eq. \eqref{SZS} holds, we can write the collective mode propagator as

\begin{eqnarray}\label{G2}
    G_2(i\nu) & = & S(Z_\eta\Omega - i\nu\mathds{1})^{-1}S^{-1}Z \\
    & = & S(\Omega -i\nu Z_\eta)^{-1}Z_\eta S^{-1}Z \\
    & = & S(\Omega -i\nu Z_\eta)^{-1}S^\dagger\,.
\end{eqnarray}
The last line is Eq. \eqref{collmodeprop} in the main text.

Let us now consider translationally invariant systems, for which $H_B$ is block-diagonal in the momentum $\q$. We will write the momentum-dependent blocks as $H_B(\q)$. We are interested in the case where $H_B(\q)$ has a zero mode at $\q = 0$, and we want to understand the behaviour of the eigenvectors contained in $S(\q)$ as a function of $\q$. Much can be learned from studying the following simple example:

\begin{equation}
    H_B(\q) = \left(\begin{matrix} 1 + \epsilon_\q^2 & 1 \\ 1 & 1+\epsilon_\q^2\end{matrix}\right)\,,
\end{equation}
where $\epsilon_{\q=0}=0$, and $\epsilon_\q\ll 1$ for all $\q$. For this example we have $Z = \sigma^z$ and $X=\sigma^x$, where $\sigma^i$ are the Pauli matrices. Working to second order in $\epsilon_\q$, we find that the eigenvalues of $ZH_B(\q)$ are given by $\omega_{\pm,\q}=\pm \sqrt{2}\epsilon_\q$. The corresponding eigenvectors, normalized to have unit Euclidean norm, are given by

\begin{eqnarray}
S(\q)_{:,+} & = & \frac{1}{\sqrt{2}-\epsilon_\q + 3\sqrt{2}\epsilon_\q^2/4}\left(\begin{matrix} 1 \\ -1 + \sqrt{2}\epsilon_\q -\epsilon_\q^2 \end{matrix}\right) \nonumber\\
S(\q)_{:,-} & = & \frac{1}{\sqrt{2}-\epsilon_\q + 3\sqrt{2}\epsilon_\q^2/4}\left(\begin{matrix} 1 - \sqrt{2}\epsilon_\q +\epsilon_\q^2  \\ -1  \end{matrix}\right)\,, \nonumber
\end{eqnarray}
where we have again only kept terms up to second order in $\epsilon_\q$. Note that as $\q\rightarrow 0$, and thus $\epsilon_\q\rightarrow 0$, the two eigenvectors $S(\q)_{:,\pm}$ become identical, reflecting the fact that the generalized eigenvalue equation $H_B(0)|v\rangle = \lambda Z|v\rangle$ has only one solution in the presence of a zero mode, and the matrix $S(\q)$ becomes degenerate. The eigenvectors also satisfy

\begin{eqnarray}
 [S^\dagger(\q)ZS(\q)]_{++}& = & \sqrt{2}\epsilon_\q \\
\left[S^\dagger(\q)ZS(\q)\right]_{--} & = & -\sqrt{2}\epsilon_\q\, .
\end{eqnarray}
These equations show that if we want to adopt the normalization of the eigenvectors such that Eq. \eqref{SZS} holds, then the Euclidean norm of the eigenvectors will diverge in the $\q\rightarrow 0$ limit as $\epsilon_\q^{-1/2}$. This is a generic feature of the generalized eigenvalue problem considered here, and is not an artefact of the simple example. To see this, let us write $ZH_B = ZU D U^\dagger$, where $U$ is the unitary matrix which diagonalizes $H_B$. If $H_B$ has a zero mode with eigenvector corresponding to the first column of $U$, denoted as $U_{:,0}$, then $U_{:,0}$ is also a zero mode of $ZH_B$, i.e. we can take $S_{:,0} = U_{:,0}$. Furthermore, as $H_B$ is particle-hole symmetric [Eq. \eqref{PHapp}], it follows that $[XU^*]_{:,0} = e^{i\alpha} U_{:,0}$, where $e^{i\alpha}$ is an unimportant phase. As $X$ anti-commutes with $Z$, the particle-hole symmetry implies that

\begin{equation}\label{SZSzero}
 [S^\dagger ZS]_{ss} = 0 \text{ if } \Omega_{ss}=\omega_s = 0 \, .
\end{equation}
This shows that the zero mode eigenvectors cannot be normalized such that Eq. \eqref{SZS} holds. Previously, we assumed a small non-zero symmetry breaking field to gap the Goldstone modes at $\q=0$. In this case, the Euclidean norm of the Goldstone mode eigenvectors at $\q=0$, normalized such that Eq. \eqref{SZS} holds, will diverge as $\omega_\q^{-1/2}$ when we take this symmetry-breaking field to zero at the end of the calculation. To avoid this divergence, we defined the rescaled wavefunctions in Eq. \eqref{rescaled} of the main text.

\section{Properties of the Goldstone mode wavefunctions and gauge fixing}\label{sec:properties}

In the main text we have introduced rescaled collective mode wavefunctions $\tilde{\varphi}^{\pm s}_\q$, whose definition we repeat here for ease of presentation:

\begin{equation}\label{apprescaled}
    \tilde{\varphi}^{\pm s}_{\q} := \sqrt{\frac{2\omega_{\q,s}aw_s N}{c_s} }\varphi_{\q}^{\pm s}\,.
\end{equation}
Here $a$ is the lattice constant, and $w_s$ a dimensionless number that will be determined below. It follows from the general discussion in the previous appendix that the components $\tilde{\varphi}^{\pm s}_{\q,\alpha\beta}(\k)$ are generically order one numbers which for fixed non-zero $\q$ do not go to zero in the thermodynamic limit. Moreover, in the thermodynamic limit $\tilde{\varphi}^{\pm s}_{\q,\alpha\beta}(\k)$ also remains finite as $\q\rightarrow 0$. As was also mentioned in the main text, we consider systems with a time-reversal symmetry $\mathcal{T}$, such that there are two degenerate collective mode wavefunctions $\tilde{\varphi}^{\pm s}_\q$ (distinguished by the $\pm$ sign) at every non-zero $\q$ which are mapped to each other by the action of $\mathcal{PT}$, where $\mathcal{P}$ is the collective-mode particle-hole symmetry.

In the main text we have used a particular gauge fixing for the collective mode wavefunctions. We now explain the details of this gauge fixing. Our main goal is to have a non-singular and smooth $\q\rightarrow 0$ limit for the collective mode wavefunctions. In particular, we want the following equation to hold:
\begin{equation}\label{lim}
\lim_{\q\rightarrow 0} \tilde{\varphi}^{\pm s}_{\q,\alpha\beta}(\k) = \tilde{Q}_{s,\alpha\beta}(\k)\,,
\end{equation}
where $\tilde{Q}_s$ is the exact $\q=0$ Goldstone zero mode defined in Eq. \eqref{zeromode}. We again repeat this definition here for ease of presentation:
\begin{equation}\label{appzeromode}
    \tilde{Q}_{s,\alpha\beta}(\k) = i\langle u_\alpha(\k)|Q_s|u_\beta(\k)\rangle \left[n_\alpha(\k)-n_\beta(\k)\right]\,.
\end{equation}
Note that right-hand side of Eq. \eqref{lim} is identical for the two different collective mode wavefunctions distinguished by $+s$ and $-s$, which implies that these two wavefunctions $\tilde{\varphi}^{\pm s}_{\q,\alpha\beta}(\k)$ obtained at non-zero $\q$ have to become identical as $\q \rightarrow 0$. This is possible because the Goldstone mode wavefunctions are \emph{not} orthogonal eigenvectors obtained from a Hermitian eigenvalue problem, but are instead obtained from the \emph{generalized} eigenvalue equation in Eq. \eqref{collmode}. Exactly at $\q =0$, there is only one Goldstone wavefunction, which corresponds to the zero mode in Eq. \eqref{appzeromode}. This happens because in the presence of zero modes, solutions to the generalized eigenvalue equation \eqref{collmode} do not form a complete basis \cite{Thouless1961}. 

As a further consistency check on Eq. \eqref{lim}, let us note that it follows from Eq. \eqref{symplnorm} that the rescaled wavefunctions satisfy
\begin{equation}\label{symplnormtilde}
\frac{1}{N}\sum_\k \sum_{i,n} \left(|\tilde{\varphi}^{\pm s}_{\q,ni}(\k)|^2 - |\tilde{\varphi}^{\pm s}_{\q,in}(\k)|^2 \right) = \pm 2\omega_{\q,s}\frac{aw_s}{c_s}\,.
\end{equation}
The exact zero mode $\tilde{Q}_s$ defined in Eq. \eqref{appzeromode} also satisfies this equation with $\omega_{\q=0,s}=0$.

In order to satisfy Eq. \eqref{lim} we have to fix both the phase and norm of the wavefunctions $\tilde{\varphi}^{\pm s}_{\q}$. The norm we fix via the dimensionless parameter $w_s$ defined in Eq. \eqref{apprescaled}. The phase can be partially fixed by using the symmetries $\mathcal{P}$ and $\mathcal{PT}$. For this we first note that the exact zero mode $\tilde{Q}_s$ is even under $\mathcal{P}$ and has eigenvalue $(-1)^{\kappa +1}$ under $\mathcal{PT}$, where $\kappa$ encodes whether the broken symmetry generator $Q_s$ is time-reversal even or odd: 
\begin{equation}
    \mathcal{T}Q_s\mathcal{T}^{-1} = (-1)^\kappa Q_s\, .
\end{equation}
Motivated by the symmetry properties of the exact zero mode, we partially fix the phase of the collective mode wavefunctions at non-zero $\q$ by requiring that the following equations hold:
\begin{eqnarray}
    \mathcal{PT}[\tilde{\varphi}^{\pm s}_\q] & = & (-1)^{\kappa +1} \tilde{\varphi}^{\mp s}_\q \label{gaugePT}\\
    \mathcal{P}[\tilde{\varphi}^{\pm s}_\q] & = & \tilde{\varphi}^{\mp s}_{-\q}\, . \label{gaugeP}
\end{eqnarray}
After this partial gauge fixing, there is a remaining phase freedom over half of the Brillouin zone for one of the two wavefunctions, say $\tilde{\varphi}_\q^{+s}$. We fix this phase by requiring $\tilde{\varphi}_\q^{+s}$ to be a continuous function of $\q$, and Eq. \eqref{lim} to hold. Note that if there is an additional inversion symmetry $\mathcal{I}$ under which $\q\rightarrow -\q$, then we can fix the remaining phase of $\tilde{\varphi}_\q^{+s}$ over half of the Brillouin zone up to a minus sign by requiring that $\tilde{\varphi}_\q^{+s}$ is either even or odd under $\mathcal{IT}$ (depending on whether $\tilde{Q}_s$ is even or odd under $\mathcal{IT}$). The requirements of continuity and a smooth $\q\rightarrow 0$ limit as in Eq. \eqref{lim} then uniquely fix the phase.

With the gauge choice in Eq. \eqref{gaugePT}, it follows that $(\tilde{\varphi}^{s}_\q \pm \tilde{\varphi}^{-s}_{\q})/2$ is an eigenstate of $\mathcal{PT}$ with eigenvalue $\pm(-1)^{\kappa + 1}$. In the limit $\q\rightarrow 0$, the symmetric combination therefore becomes the unique zero mode $\tilde{Q}_s$, which has $\mathcal{PT}$ eigenvalue $(-1)^{\kappa +1}$. The anti-symmetric combination becomes a zero mode with $\mathcal{PT}$ eigenvalue $-(-1)^{\kappa +1}$, which must be the zero vector. From this it follows that with the gauge choice in Eq. \eqref{gaugePT}, equation \eqref{lim} indeed holds.

\end{appendix}

\bibliography{bib}
\end{document}